\documentclass[preprint]{aastex}
\usepackage{apjfonts}[11pt]

\begin{document}
\title{Global Star Formation Rate Density over $0.7<z<1.9$}

\author{Hyunjin Shim\altaffilmark{1}, James Colbert\altaffilmark{2},
 Harry Teplitz\altaffilmark{2}, Alaina Henry\altaffilmark{3},
 Mattew Malkan\altaffilmark{3}, Patrick McCarthy\altaffilmark{4},
 \& Lin Yan\altaffilmark{2} }

\altaffiltext{1}{Department of Physics \& Astronomy, FPRD,
 Seoul National University, Seoul, Korea ; hjshim@astro.snu.ac.kr}
\altaffiltext{2}{Spitzer Science Center, California Institute of Technology,
 Pasadena, CA}
\altaffiltext{3}{Department of Physics \& Astronomy, University of California, 
 Los Angeles, CA 90095-1547; malkan@astro.ucla.edu}
\altaffiltext{4}{Observatories of the Carnegie Institution of Washington, 
 813 Santa Barbara Street, Pasadena, CA}

  \begin{abstract}
  We determine the global star formation rate density at $0.7<z<1.9$
  using emission-line selected galaxies identified in Hubble Space Telescope
  Near Infrared Camera and Multi-Object Spectrograph (HST-NICMOS) grism
  spectroscopy observations. Observing in pure parallel mode throughout HST
  Cycles 12 and 13, our survey covers $\sim104$ arcmin$^2$ from which we select
  80 galaxies with likely redshifted H$\alpha$ emission lines. In several cases,
  a somewhat weaker [OIII] doublet emission is also detected.
  The H$\alpha$ luminosity range of 
  the emission-line galaxy sample is
  $4.4\times10^{41} < L (\mbox{H}\alpha) < 1.5\times10^{43}$ erg s$^{-1}$. 
  In this range, the luminosity function 
  is well described by a Schechter function with $\phi^* = (4.24\pm3.55)\times10^{-3}$
  Mpc$^{-3}$, $L^* = (2.88\pm1.58)\times10^{42}$ erg s$^{-1}$, and $\alpha = -1.39\pm0.43$. 
  We derive a volume-averaged star formation rate density of 
  $0.138\pm0.058 \mbox{M}_{\odot} \mbox{yr}^{-1} \mbox{Mpc}^{-3}$ at $z=1.4$
  without an extinction correction. 
  Subdividing the redshift range, we find star formation rate densities of 
  $0.088\pm0.056 \mbox{M}_{\odot} \mbox{yr}^{-1} \mbox{Mpc}^{-3}$ at $z=1.1$ and
  $0.265\pm0.174 \mbox{M}_{\odot} \mbox{yr}^{-1} \mbox{Mpc}^{-3}$ at $z=1.6$.
  The overall star formation rate density is consistent with previous
  studies using H$\alpha$ when the same average extinction correction is applied,
  confirming that the cosmic peak of star formation occurs at $z>1.5$.
  \end{abstract}


\section{INTRODUCTION} 

 The evolution of cosmic star formation rate density (SFRD) and 
 the stellar mass function are two key components required to describe
 galaxy evolution, representing current and past star formation activities
 respectively. Measurements of the SFR density at different redshifts
 (e.g., Madau et al. 1998; Steidel et al. 1999; Arnouts et al. 2005;
 Schiminovich et al. 2005; Bouwens et al. 2006, 2007; Ly et al. 2007) 
 suggest that the global SFR density peaks at $1<z<3$. The study of the
 build-up of stellar mass density (e.g., Dickinson et al. 2003; 
 Rudnick et al. 2003; Glazebrook et al. 2004; Fontana et al. 2004)
 also indicates that the redshift range $1<z<3$ is the phase of 
 massive galaxy formation, being the epoch of the strongest 
 star formation. Most studies are in overall agreement that 
 star formation decreases by a factor of 10 to 20 from $z\sim1$
 to $z=0$.  

 While the redshift range $1<z<2$ is expected as the epoch of the
 strongest star formation, measurements of the SFR density over 
 this redshift range are uncertain. First, most of the significant
 spectral features useful for redshift identification are in the near-infrared
 (0.7--2$\mu$m) at $z>1$, so few galaxies at $z\sim1$--2 have
 spectroscopic redshifts.  
 Second, the commonly used rest-frame ultraviolet or mid-infrared
 selections can be severely biased toward relatively unobscured or 
 obscured star-forming galaxy populations. Therefore we performed 
 a NIR spectroscopic survey using redshifted H$\alpha$ emission lines
 at $z>1$, to select star-forming galaxies at the corresponding redshifts.

 H$\alpha$ is known to be a robust measure of star formation, 
 which is less affected by dust extinction compared to UV continuum 
 (e.g., see review of Kennicutt 1998).
 At redshift $1<z<2$, grism spectroscopy using
 Near Infrared Camera and Multi-Object Spectrograph (NICMOS) onboard
 Hubble Space Telescope (HST) offers a unique tool to sample
 H$\alpha$-selected star-forming galaxies. The first results of
 a NICMOS grism parallel survey were published using HST Cycle 7
 data (McCarthy et al. 1999). 
 Operating in pure parallel mode, the survey identified 33 
 emission-line galaxies over $\sim85$ arcmin$^2$ of randomly selected 
 fields. The H$\alpha$ luminosity function at $z=0.7$--1.9 was
 derived using the identified emission-line galaxies (Yan et al. 1999). 
 Hopkins et al.(2000) investigate the faint-end of the H$\alpha$
 luminosity function in more detail by performing deeper pointed grism
 observations. This NICMOS parallel grism survey resumed after the
 installation of the NICMOS cryocooler and continued until 2005,
 including hundreds of observations throughout Cycles 12 and 13. 
 This extensive amount of new data enables the construction of a new,
 more robust H$\alpha$ luminosity function at $z=1$--2.
 
 In this paper, we present the H$\alpha$ luminosity function at 
 $0.7<z<1.9$. Possible evolution of the luminosity function between 
 redshift 1 and 2 is also investigated, by constructing luminosity functions
 at $0.7<z<1.4$ and $1.4<z<1.9$ separately. The global star formation
 rate density inferred by the H$\alpha$ luminosity function is compared
 with the values from previous studies, in context of the galaxy evolution. 
 Throughout this paper, we use a cosmology with
 $\Omega_{M}=0.3$, $\Omega_{\Lambda}=0.7$,
 and $H_0=71\mbox{km} \mbox{s}^{-1} \mbox{Mpc}^{-1}$.

\section{DATA} 

 \subsection{Observations}

 All of the data presented in this study have been obtained using
 camera 3 of NICMOS onboard HST, taken in pure parallel mode with 
 the G141 grism and the broad-band F110W/F160W filters. The original 
 NICMOS camera 3 image is $256\times256$ pixel array with a pixel scale of
 $0.2\arcsec$ pixel$^{-1}$, 
 providing a field of view of $51.2\arcsec\times51.2\arcsec$
 ($\sim0.75$ arcmin$^2$).
 The observations were performed between October 2003 and July 2004
 (Cycle12), July 2004 and January 2005 (Cycle13). 
 The fields are randomly selected, approximately 10$\arcmin$ apart from 
 the coordinates of the prime observation. The total exposure times 
 for different fields vary from 768 seconds to
 48000 seconds, while typical integration times ranged from $\sim2000$ to
 $\sim30000$ seconds. Every observed field has at least
 three dithered frames. To identify objects that are responsible for
 the spectra in slitless grism images, we obtained F160W ($H$-band)
 direct images before or after taking the grism images.
 For all Cycle 13 and several Cycle 12 
 fields, we also obtained F110W ($J$-band) direct images
 (For the information about the reduction and analysis of F110W and F160W
 direct images, see Henry et al. 2007, 2008).

 The G141 grism covers a wavelength range of 1.1 to 1.9$\mu$m,
 with mean dispersion of $8\times10^{-3}\mu$m pixel$^{-1}$.
 The resolving power $R$ is a function of the observing condition,
 including the variation of Point Spread Function (PSF) due to the 
 changes in the optical telescope assembly and longer term changes in
 the internal structure (See McCarthy et al. 1999 for details). According
 to the previous observations, the nominal resolution is low: 
 $R\sim100$--200 (Noll et al. 2004). Thus, most of the lines are 
 unresolved in this study. Assuming H$\alpha$ emission line redshifted 
 to $1<z<2$ and $R=100$ at $\lambda=1.5\mu$m, the smallest detectable
 rest-frame intrinsic equivalent width is 50--75$\mbox{\AA}$, although
 there is some additional uncertainty resulting from the unknown relative 
 strength of the [NII] to the H$\alpha$ line.

 \subsection{Image Reduction}

 The data reduction follows similar steps to previous studies
 using NICMOS grism data (e.g., McCarthy et al. 1999; Hopkins et al. 2000),
 including high background subtraction, one-dimensional spectra extraction,
 and wavelength/flux calibration.  
 We start from the calibrated output \texttt{*cal.fits} files from
 the HST archive, which are corrected for bias and dark current removal,
 linearization, and cosmic ray rejection. Flat-fielding is not included 
 in this stage but is done after the extraction of one-dimensional
 spectrum, since the flat-field strongly depends on the wavelengths
 in case of NICMOS grism.
 For images taken during the South Atlantic Anomaly, we apply SAA correction using 
 \texttt{saa\_clean}\footnote{http://www.stsci.edu/hst/nicmos/tools/post\_SAA\_tools.html}.
 After the SAA correction and the correction for any remaining differences in
 bias levels between each pedestal quarters, the main part of the data reduction
 is the removal of the high sky background at near-infrared wavelengths. 
 We use two different methods to make the sky frame that will be subtracted, 
 and select the better method for each case: i) median-combining all image
 frames taken and ii) median-combining only those image frames taken at close 
 dates with the image frame that needs sky subtraction. For several fields
 taken during the period in which the sky background changes rapidly, the 
 highly uneven background is not removed completely with the first method. 
 In such cases, we use the second method to make a sky frame. At lease 9 frames 
 taken at close dates were used to construct the sky frame, to prevent the
 increase in background uncertainty. Once the sky frame for each image 
 is determined, the sky frame is subtracted from the observed images. 
 Note that the construction of sky frames using image frames taken at close
 dates is newly introduced while previous studies used only one sky frame 
 constructed by median-combination of all image frames (e.g., McCarthy et al.
 1999).

 After the sky subtraction, we grouped the frames according to their
 139 unique parallel fields and registered each group onto one plane. 
 In order to measure the offsets between the dithered frames, we shift
 each frame by a series of $\Delta x$, $\Delta y$ ($\pm$0.1 pixel)
 shift values, subtract the shifted one from the reference frame, and find
 the shifts that minimize the difference through the iteration.
 The final shifts in x/y directions are less than 3 pixels in general.
 During this process, the bad pixels and hot pixels are masked out.
 For bad pixel masks, we combine the permanent bad pixel mask for the 
 NICMOS camera 3 and the data quality flag image associated with the
 NICMOS raw data cube. Some ``warm'' pixels, which are missed in bad pixel 
 masks are identified by eye and added in the final mask for correction. 
 Final cleaned, sky-subtracted, coordinate-registered frames are added to 
 construct the reduced mosaic two-dimensional image for each pointing.

 From the reduced two-dimensional spectra image, 
 we extract one-dimensional spectrum and perform flux/wavelength calibration
 using the \texttt{NICMOSlook}
 software\footnote{http://www.stecf.org/instruments/NICMOSgrism/nicmoslook/nicmoslook/index.html}
 developed by STEC-F (Freudling 1999). 
 We first identify the position of the emission-line galaxy in
 the direct image (F160W) through visual inspection on the two-dimensional 
 spectra image. From the interactively determined position and size of the object,
 we define the extraction aperture and the background region considering the offset
 between the grism and the direct image, and extract the spectrum.

 We apply a correction for the wavelength dependent pixel response 
 to the extracted spectrum using an inverse sensitivity curve.
 For wavelength calibration, we first extract the spectrum of 
 a bright point source in each grism image before extracting the
 spectrum of any emission-line galaxies. The bright point sources
 reproduce the significant cutoff in short/long limits (1$\mu$m, 1.9$\mu$m)
 in inverse sensitivity curve with a high $S/N$, because stellar spectra are flat 
 in the G141 bandpass. 
 Final wavelength calibration for the spectrum of
 emission-line galaxy is done by adjusting slight offsets between 
 the spectrum and the overlaid sensitivity curve. 
 Note that sometimes the wavelength calibration varies for different
 locations in a camera field of view. The uncertainty in wavelength
 calibration due to the remaining distortion effect is about 
 $\sim0.02\mu$m, i.e., in general there is a systematic redshift determination
 error of $\Delta z\sim0.03$.
 Also, there are a few cases where the emission-line 
 object lies near the edge of the image. In these cases, the counterpart of the
 object is not found in the associated direct image, which causes a large 
 uncertainty in wavelength calibration up to $\sim0.1\mu$m. The redshift
 uncertainty for an object lying at the image edge is
 $\Delta z\sim0.15$. 
 Finally, the extracted one-dimensional spectrum is flux-calibrated by 
 dividing the pixels values ([DN/s]) by the G141 grism inverse sensitivity curve 
 ([DN/s]/Jy).

 Our reduction method is comparable with that used in Hubble Legacy Archive
 (Freudling et al. 2008). The consistency between our spectra and the reduced
 spectra in Hubble Legacy Archive\footnote{http://hla.stecf.org} confirms 
 the existence of emission lines for the sample galaxies, although we find
 matches for only a few objects which are the most luminous. Most of
 our emission line galaxies are relatively faint, and are therefore missed by
 the less rigorous reduction in the Hubble Legacy Archive. 
 
\subsection{Area Coverage and Depth}

 This work is an extension of the previous NICMOS grism survey for
 emission-line galaxies from the Cycle 7 parallel observation program
 (McCarthy et al. 1999), performed during Cycle 12 and Cycle 13.
 In Cycle 12, our survey targeted
 130 different coordinates with different exposure times. 
 From the original 130 fields, 26 are excluded in the
 final analysis since the fields are either too crowded 
 (i.e., the stellar densities are over $\sim50$ arcmin$^{-2}$, 
 M31/SMC fields), have high galactic foreground extinction
 (Taurus Molecular Cloud fields), or are damaged by latents produced by 
 very bright objects observed just before the image exposure.
 In Cycle 13, we excluded 10 fields out of the 45 fields initially targeted
 for the same reasons. Therefore the total survey area is $\sim104$ arcmin$^2$,
 covering 139 different fields randomly distributed over the sky. 

\begin{figure*}
\plottwo{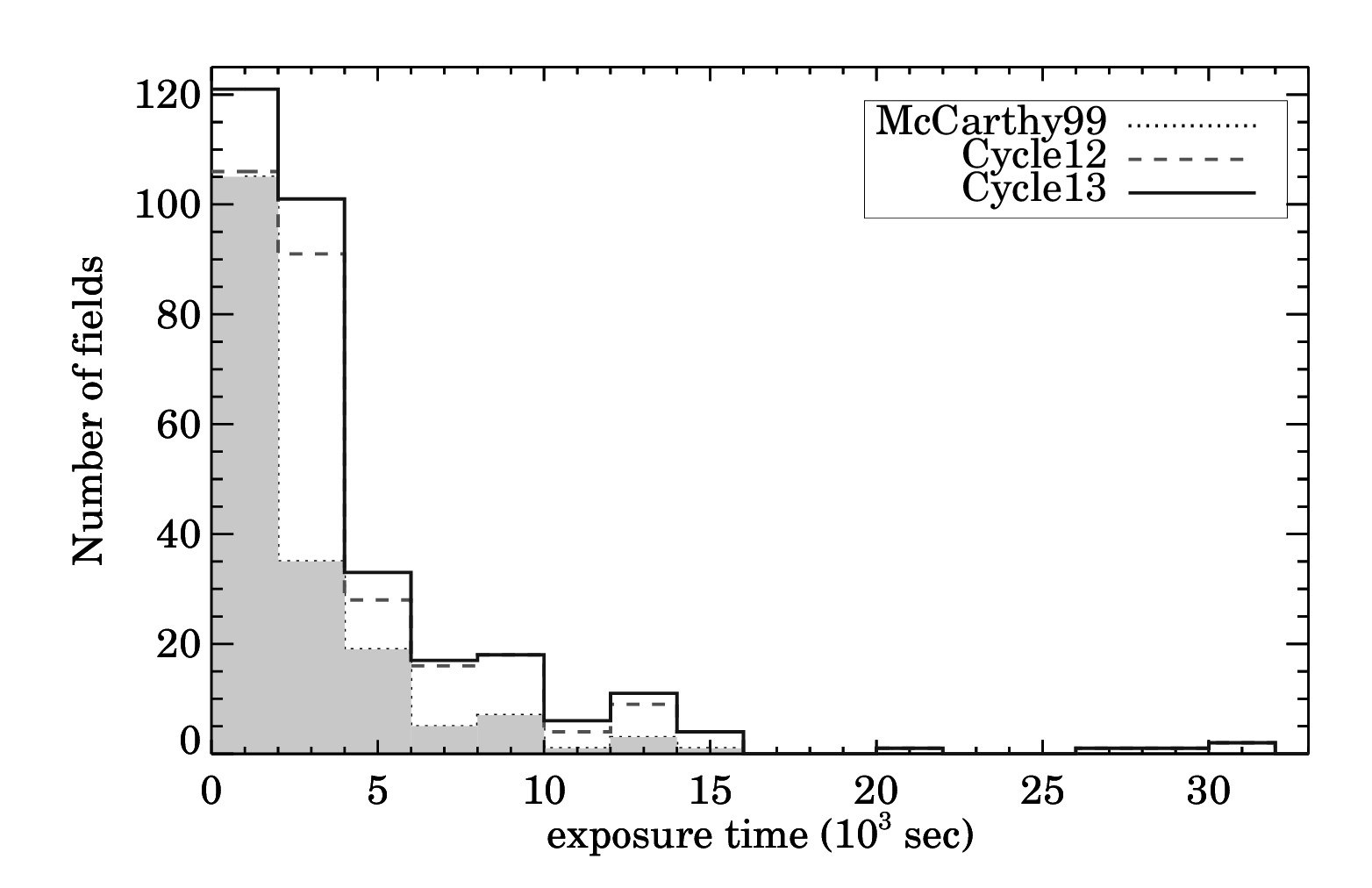}{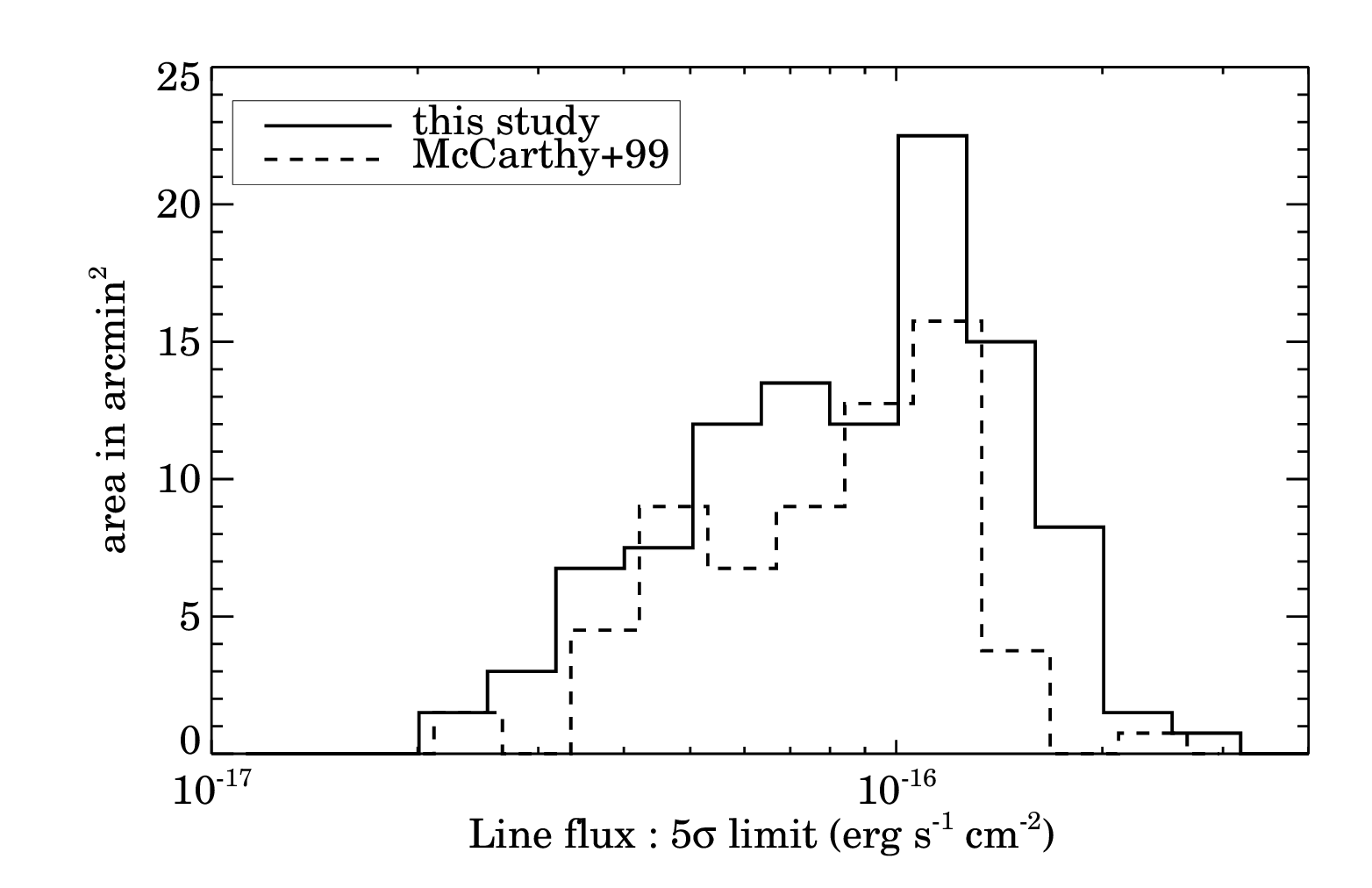}
\caption{\label{fig:data} (\textit{Top}):
 The distribution of exposure times of the observed fields in
 Cycles 7, 12, and 13. The shaded histogram
 with \textit{dotted} lines is the data presented in McCarthy et al.
 (1999). The open histogram with \textit{dashed}/\textit{solid}
 line is the accumulated distribution when Cycles 12/13 data are
 added to the existing data. That is, the blue \textit{dashed}
 line indicates the sum of Cycle 7 and Cycle 12 data, while
 the red \textit{solid} line indicates the sum of Cycle 7,
 Cycle 12, and Cycle 13 data.
 (\textit{Bottom}) :
 The histogram of $5\sigma$ line flux limits, within a 4-pixel
 aperture of the NICMOS grism data. Each pointing covers
 0.75 arcmin$^2$.
 The line flux limits are compared to those from McCarthy et al.
 (1999), which is drawn with a \textit{dashed} line.
 }
\end{figure*}

 We compare the exposure times of these 139 useable parallel fields taken during
 Cycles 12 and 13 with the earlier Cycle 7 survey in Figure \ref{fig:data}a.
 Although the number of the deepest exposures is not significantly increased
 in Cycles 12 and 13, the number of total pointings is nearly double those 
 from the Cycle 7 (Cycle 7 data comprises 85 pointings over $\sim65$ arcmin$^2$; 
 McCarthy et al. 1999). In particular,  the number of pointings with medium 
 exposure times (2000--10000 seconds) have increased substantially. 

 We illustrate the distribution of $5\sigma$ line flux limits in Figure 
 \ref{fig:data}b. The rms noise in spectra-free regions is measured over
 a 4-pixel aperture for extracting one-dimensional spectrum. Thus, this
 ``line flux limit'' reflects the flux limit of a line added to the
 underlying continuum. As is expected from the exposure time comparison
 between Cycle 7 and Cycle 12/13 (Figure \ref{fig:data}a), the line flux
 limits distribution of our data shows similar trend with that of Cycle 7. 
 Though it is true that the fields with longer exposure time have fainter 
 line flux limits, the depth of an image is not necessarily a simple
 function of the exposure time of an image. Instead, the flux limit is
 much more dependent on the flatness of the background, i.e., non-Poisson
 noise caused by imperfect background subtraction. 
 Therefore, below $5\sigma$ flux limits, we don't reliably identify
 emission lines because significant residuals remain from dark subtraction 
 and other large data artifacts.

\section{EMISSION-LINE GALAXIES}

  \subsection{Identification}

\begin{figure*}
 \plotone{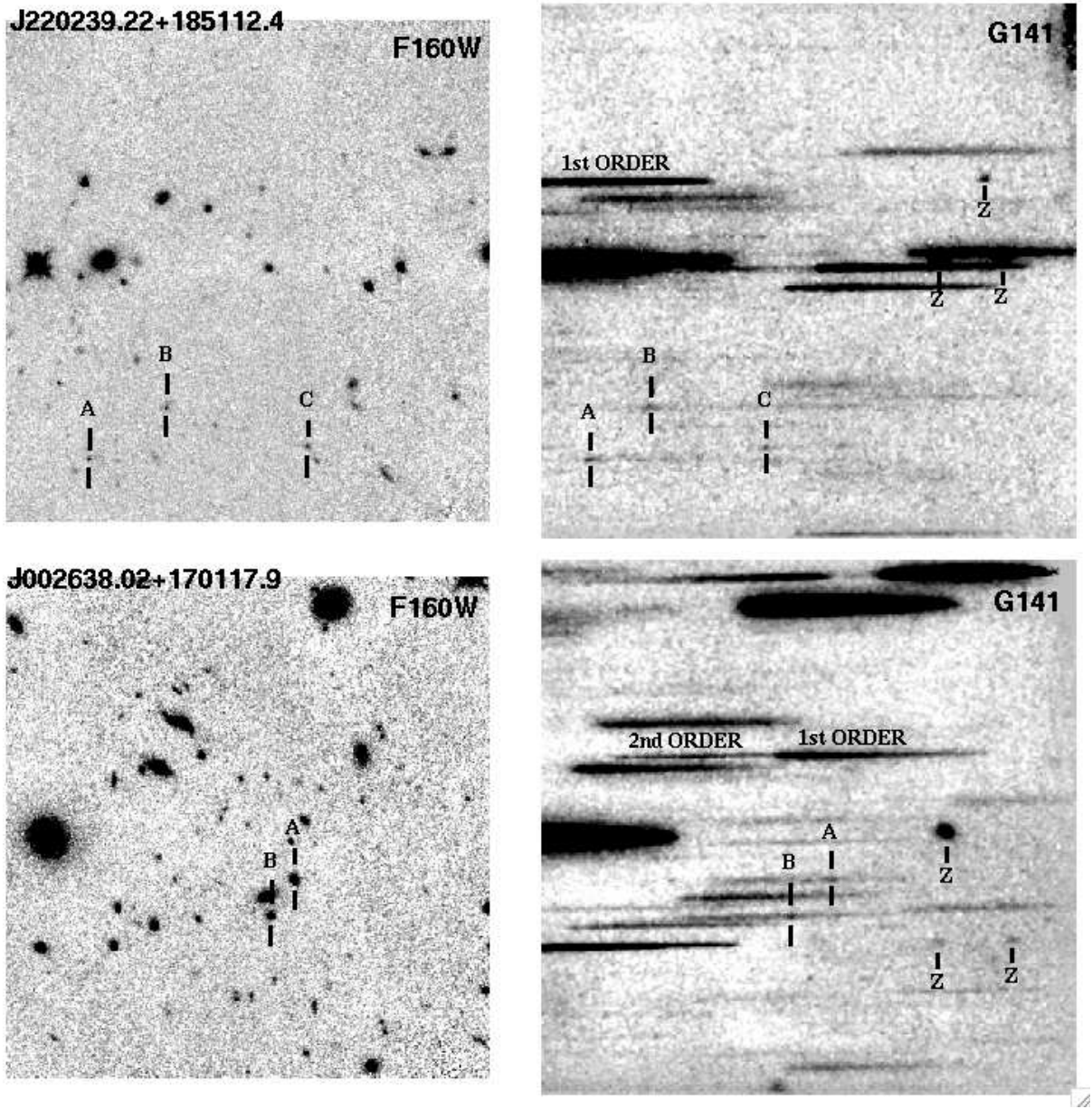}
 \caption{ \label{fig:identify}
 Typical pairs of direct (F160W) and grism (G141) images of our NICMOS
 parallel survey. Each field is $51.2\arcsec\times51.2\arcsec$.
 In the grism two-dimensional image on the right, we mark the
 zero-order image (Z), the first and second order spectrum
 (1st order / 2nd order), and the emission lines. The objects
 producing the emission lines are also marked as A, B, or C
 in the direct image.  }
\end{figure*}

 We identify the emission-line galaxy candidates on the two-dimensional
 spectra image through visual inspection, prior to the extraction of a
 one-dimensional spectrum using \texttt{NICMOSlook} (Section 2.2). 
 We carefully compare the grism images and the direct images, identifying
 the galaxies in the direct images that are responsible for
 the emission-line in the spectra. We cross-check this method by 
 identifying the same emission-line galaxies among multiple
 different authors.
 Some major obstacles in the identification of the emission-lines are 
 the existence of zero-order images, the remaining background patterns,
 and the occasional image artifact. The zero-order image appears at
 a location $\sim27\arcsec$ apart from the end of the first order spectrum,
 so it can be recognized in most cases through the inspection of
 the first-order spectrum or the inspection of direct images.
 However in the middle portion of the detector, it is difficult
 to tell whether the point-like feature is a zero-order image or
 a strong emission-line with faint continuum. 
 Figure \ref{fig:identify} shows two typical pairs of direct and grism
 images in our data and the identified emission-line galaxies. Zero-order
 images, the first and the second order spectrum, and the identified
 emission-line features are marked.

 Over the $\sim104$ arcmin$^2$ surveyed in Cycle 12 and Cycle 13, 
 we identify 80 emission-line galaxies. As we mentioned in Section 2.3,
 we only classify an emission-line as real if the line is significant 
 at $>5\sigma$ levels. Table \ref{tab:objects} presents
 the coordinates, redshifts, line fluxes, observed equivalent widths,
 and available photometry of all 80 emission-line galaxies. 
 The emission-line galaxies are distributed over 53 different fields,
 with 23 fields containing more than one emission-line galaxy.

 The most likely candidates for the identified emission lines in our 
 grism survey are hydrogen lines (H$\alpha$, H$\beta$) and oxygen lines
 ([OII] 3727$\mbox{\AA}$, [OIII] 5007$\mbox{\AA}$) redshifted to $z>0.7$.
 Because of the broad wavelength coverage of G141 (1.1--1.9$\mu$m), we 
 would expect to see both [OIII] and H$\alpha$ in galaxies at $1.2<z<1.9$
 and both [OII] and [OIII] in galaxies at $1.95<z<2.8$. 
 Only a single line is predicted for H$\alpha$ at $z<1.2$ and [OII] at $z>2.8$. 
 Except for ten possible cases discussed in Section 3.3, we do not 
 convincingly detect more than one emission line in most of the galaxies. 
 That is, the derivation of redshift depends on only a single strong
 emission line in most cases. We believe that most of these single
 lines are H$\alpha$ considering their large equivalent widths.

\begin{figure*}
   \epsscale{1.0}\plotone{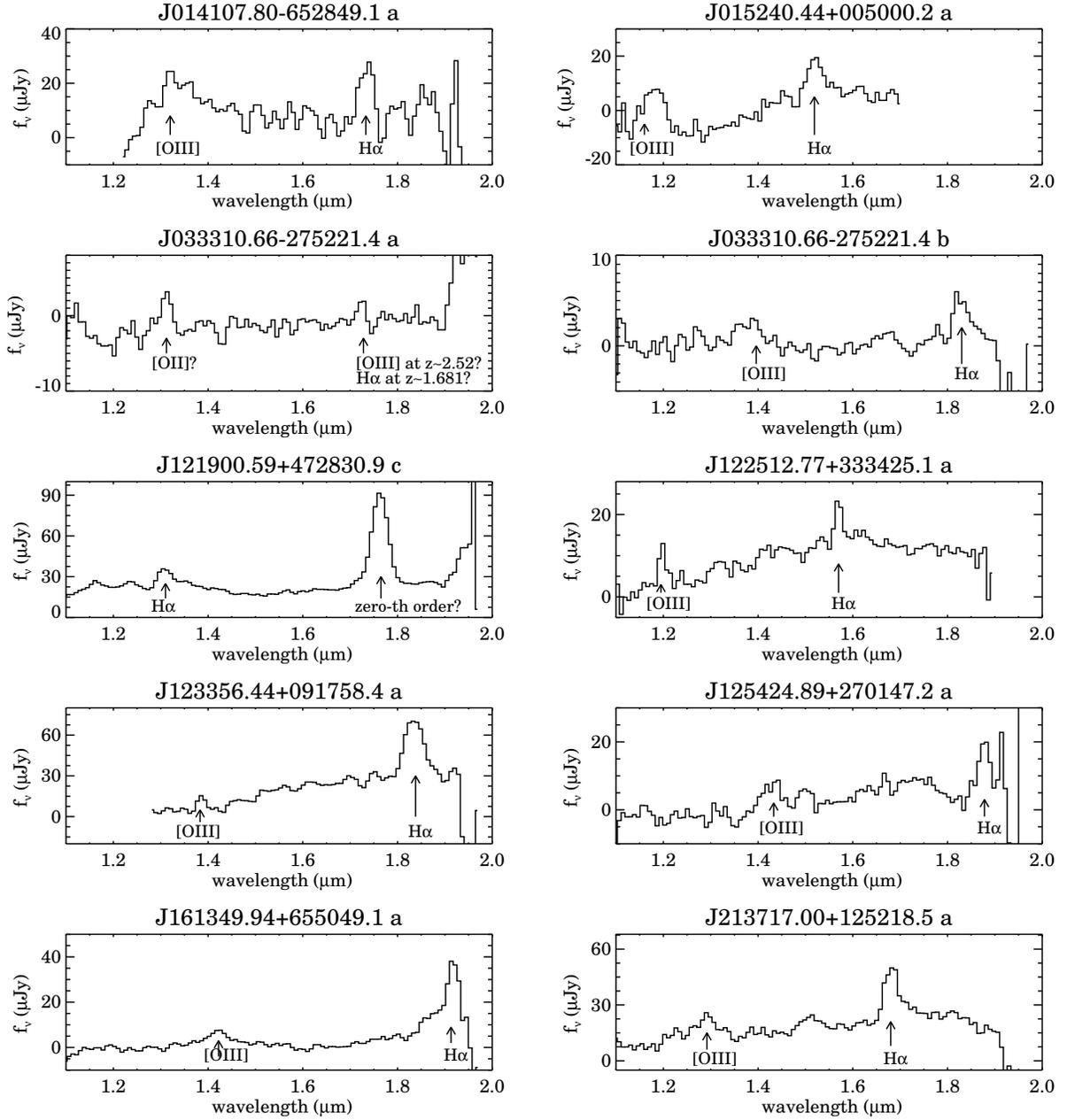}
   \caption{\label{fig:comment} One-dimensional spectra of the
   galaxies which appear to have more than one emission line.
   In most cases, the emission lines are redshifted H$\alpha$ and
   [OIII] 5007$\mbox{\AA}$ (see Section 3.3 for details).
   }
\end{figure*}

 We can estimate the possibilities of the emission lines being emission
 lines other than H$\alpha$ using the expected equivalent widths of 
 the lines, since our identification is limited to emission lines with
 EW (rest-frame) $> 40$--50$\mbox{\AA}$. First, the possibility 
 of H$\beta$ line is removed since the average equivalent widths for
 H$\beta$ in star-forming galaxies are known to be relatively small
 (5--$10\mbox{\AA}$, Brinchmann et al. 2004). The next strongest line 
 after H$\alpha$ in terms of equivalent width is [OII] 3727$\mbox{\AA}$
 line. Identifying the emission line as [OII] 3727$\mbox{\AA}$ requires
 the redshift of the object to be $1.95<z<4$. Considering the typical
 magnitude of our emission-line galaxies, the $M_V$ should be $\sim-24$
 mag if the galaxy is at $z\sim3$. The estimated number of $z\sim3$ 
 galaxies with $M_V<-24$ over our survey volume is less than two according
 to the V-band luminosity function of $z\sim3$ galaxies (Shapley et al. 
 2001), thus the possible contamination rate by [OII] lines at $z\sim3$
 is less than 3 percents (2/80).  Furthermore, most [OII] contaminants
 will be in the $1.95<z<2.8$ range, where [OIII] 5007$\mbox{\AA}$ line
 should also appear in our spectra. 
 We suggest this possibility for one object, J033310.66-275221.4a
 presented in Figure \ref{fig:comment} in Section 3.3. This object is also
 included in Table \ref{tab:objects} since the possibility of the line
 being [OII] instead of H$\alpha$ is still uncertain. 
 Finally, the last possibility for contamination is the [OIII] 5007$\mbox{\AA}$ line.
 Yet as we mentioned in the previous paragraph, [OIII] 5007$\mbox{\AA}$
 emission line is accompanied with either H$\alpha$ or [OII] 3727$\mbox{\AA}$
 at $1.2<z<2.8$. Therefore the possibility of [OIII] being the 
 single, strongest emission line of the galaxy is very low. 

 Besides these strong candidates, lines with longer wavelengths are
 also possible, including HeII $10830\mbox{\AA}$.
 However, these are not likely due to
 their weakness and the lack of accompanying nearby lines like [SIII] at
 $9069, 9545\mbox{\AA}$.
 The follow-up optical spectroscopy targeting 14 emission-line galaxies
 presented in McCarthy et al.(1999) revealed that 9 of the emission-line galaxies
 are truly at $z=1$--2, supporting the identified emission lines
 through NICMOS grism spectroscopy are mostly redshifted H$\alpha$
 (Hicks et al. 2002). The remaining 5 sources had no identifiable
 emission lines and were therefore unconfirmed, but not ruled out.
 In conclusion, we find  that contamination from other emission lines
 is small ($\le 5$\%). Nonetheless, we include this source of uncertainty
 in the derivation of the H$\alpha$ luminosity function (Section 4). 
 
 In addition to mis-identification of emission lines, there is also a 
 possibility of contamination by H$\alpha$ emission from AGN. 
 We estimated the fraction of these 
 contaminants using AGN luminosity function at similar redshifts and
 magnitudes. At $z\simeq1$, the number density of Type-1 AGN with 
 $\langle M_{B}\rangle\sim-23$ mag, corresponding to the median 
 magnitude of our emission-line galaxies, is $\sim10^{-5}$ Mpc$^{-3}$
 (Bongiorno et al. 2007). Our effective survey volume is 
 $\sim3.14\times10^5$ Mpc$^3$ at this redshift range, which would 
 indicate 3--4 AGNs in our survey. This is actually the lower limit
 on the AGN number density in our data, since for a given $H$-band 
 continuum magnitudes, AGN candidates are likely to be brighter and 
 thus easier to be included in the grism survey sample 
 compared to normal star-forming galaxies, due to 
 the larger intrinsic equivalent width of the H$\alpha$ line.
 Using the follow-up optical spectroscopy of the objects selected in 
 the previous NICMOS grism survey, Hicks et al.(2002) suggested a new
 diagnostic for Seyfert 1 galaxies with L(H$\alpha$) and H$\alpha$
 equivalent width. We have five objects within the conservative cut 
 of log(L$_{\mbox{H}\alpha}$)$>42.6$ erg s$^{-1}$ and
 EW(H$\alpha$)$>100\mbox{\AA}$, while we have five more objects with
 EW(H$\alpha$)$>100\mbox{\AA}$ and the luminosities of 
 $42.5<$log(L$_{\mbox{H}\alpha}$)$<42.6$ erg s$^{-1}$. We assume these 
 numbers represent the possible uncertainties caused by AGN contamination.
  
 To summarize, i) all the galaxies in the brightest bin
 (log(L$_{\mbox{H}\alpha}$)$>42.8$ erg s$^{-1}$) can be considered 
 AGNs according to the criteria of Hicks et al.(2002). ii) $\sim40$\%
 of the galaxies in the bin of $42.5<$log(L$_{\mbox{H}\alpha}$)$<42.8$
 erg s$^{-1}$ are possible AGNs with large equivalent widths. These
 uncertainties are included in Table \ref{tab:LF_val}, and in the derivation
 of the H$\alpha$ luminosity function. 
 
 \subsection{H$\alpha$-Derived Star Formation Rates} 

 Since H$\alpha$ and [NII] 6583$\mbox{\AA}$, 6548$\mbox{\AA}$ are not 
 deblended at the resolution of NICMOS grism, we correct the derived
 H$\alpha$ luminosity for [NII] contribution when we use H$\alpha$ 
 luminosity as measures of star formation rate. The flux ratio used 
 in correction is [NII] 6583/H$\alpha = 0.3$
 and [NII] 6583/[NII] 6548 $= 3$ (Gallego et al. 1997). The
 H$\alpha$ luminosities used throughout this paper are derived
 from this corrected H$\alpha$ flux, derived using the following formula:
 $f_c (\mbox{H}\alpha) = 0.71\times f(\mbox{H}\alpha + [\mbox{NII}])$.
 The star formation rates are derived from the corrected H$\alpha$
 luminosity using the formula of Kennicutt (1998), which assume
 a Salpeter IMF between 0.1--100 M$_{\odot}$. Note that the star formation
 rate in Table \ref{tab:objects} is not corrected for dust extinction.

\begin{figure*}
 \epsscale{1.1}\plottwo{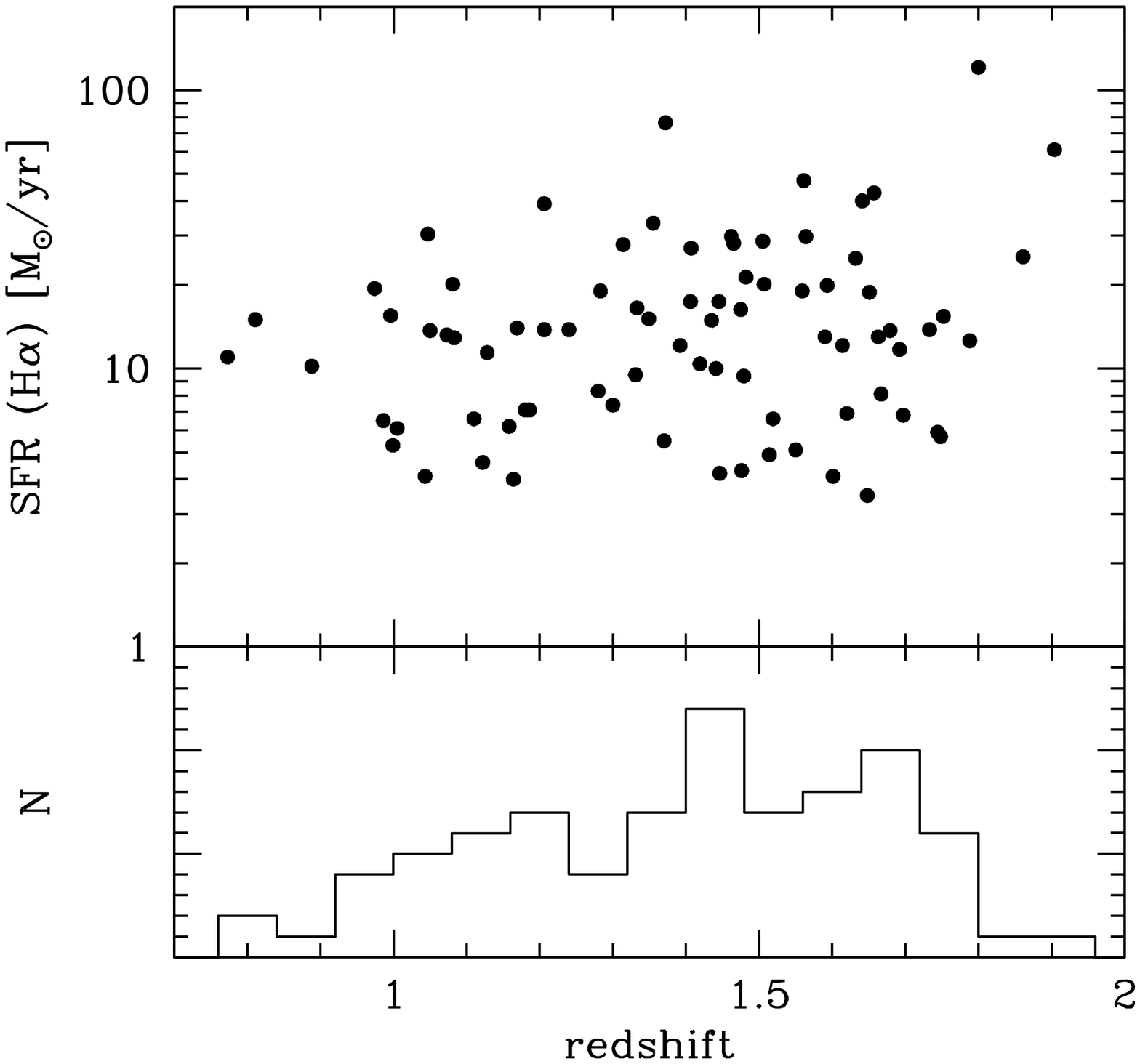}{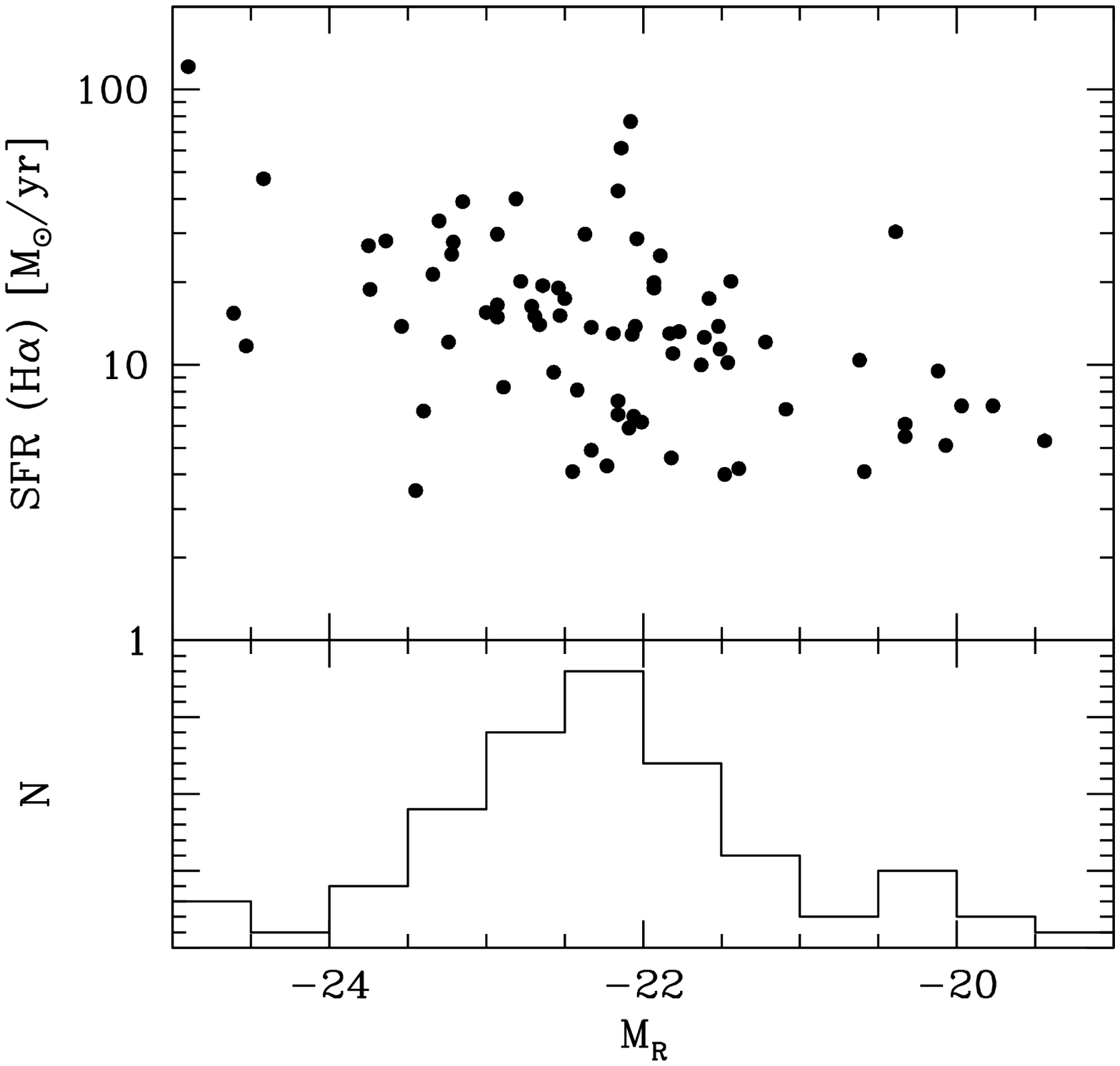}
 \caption{ \label{fig:sfrdist}
 (\textit{Left}): Star formation rates of the emission-line galaxies
 as a function of redshift. The identified emission-line galaxies
 are distributed over $0.7<z<1.9$, with no significant redshift
 peak. (\textit{Right}): Star formation rates of the emission-line
 galaxies as a function of absolute magnitudes $M_R$.
 $M_R$ magnitudes are derived from the observed $H$-band (F160W) 
 magnitudes.}
\end{figure*}

 In Figure \ref{fig:sfrdist},
 we show the distribution of H$\alpha$-derived star formation rates
 as a function of redshift and M$_R$. 
 The H$\alpha$-inferred star formation rates
 of the identified emission-line galaxies are
 2--200 $\mbox{M}_{\odot}\mbox{yr}^{-1}$,
 comparable to or larger than the UV-estimated star formation rates of 
 typical $z\sim3$ Lyman break galaxies
 with relatively little extinction (Shapley et al. 2001).
 We do not apply any dust extinction correction for the star formation
 rate of individual galaxies or our derivation of H$\alpha$ luminosity
 function, although 
 $\langle A_{V} \rangle \sim1$ mag is expected for H$\alpha$-selected
 star-forming galaxies (e.g., Kennicutt 1992). 
 We only use this H$\alpha$ extinction 
 correction for our estimates of star formation rate density 
 evolution (see Section 5). 
 
 The redshifts of our sample galaxies span the range of $0.7<z<1.9$,
 with a steep decrease at the low-redshift ($z<0.9$) and
 high-redshift ends ($z>1.8$). This is expected from the sharp end of
 the G141 grism response curve. The median redshift is
 $\langle z\rangle=1.4$. The inhomogeneous redshift distribution is
 considered in the derivation of the luminosity function,
 but the effect is small.

 Figure \ref{fig:sfrdist}b illustrates the median absolute magnitude 
 of $M_{R}\sim-22.5$ for our sample galaxies. The magnitude is 
 comparable with $M_{B}^*=-22.8$ for galaxies at
 $z=1$--1.2 (Ryan et al. 2007). Therefore our galaxies have typical
 luminosities around $M^*$, across the entire
 redshift range. Despite considerable scatter, $M_{R}$ can be used
 as stellar mass indicator for a galaxy. Thus, the higher SFR for
 the brighter galaxies may indicate that between $1<z<2$, the star
 formation is higher in the more massive galaxies in our sample.

 \subsection{Spectra of Possible Double-Line Objects}

 We discussed in Section 3.1, the validity of identifying
 the single emission lines as H$\alpha$. 
 If more than one emission line (i.e., two emission lines) exist, 
 in nearly every case we identify them to be redshifted H$\alpha$
 and [OIII] 5007$\mbox{\AA}$ based on the wavelength ratio between
 two emission lines.
 Using the known [OIII]/H$\alpha$ EW ratios of
 star-forming galaxies at $z\sim0.7$, we predict the number of
 galaxies showing both H$\alpha$ and [OIII].
 For a typical [OIII]/H$\alpha$ EW ratio of $\sim0.4$\footnote{
 This is the ratio calculated for the NICMOS grism emission-line
 objects using the follow-up optical spectroscopy
 (McCarthy et al. 1999; Hicks et al. 2002).
 The ratio is also comparable with the 
 [OIII]/H$\alpha$ EW ratio of emission-line galaxies in HST/ACS
 grism parallel survey (Drozdovsky et al. 2005) at $0.5<z<0.7$. }
 and our detection limit of EW(emission line)$>40$--50$\mbox{\AA}$,
 the H$\alpha$ equivalent width required for
 [OIII] 5007$\mbox{\AA}$ line detection is $>120\mbox{\AA}$.
 Additionally, in order to show both [OIII]/H$\alpha$ lines in grism
 spectrum, the redshift of the object should be $1.2<z<1.9$ which
 corresponds to $\sim60$\% of the total survey volume.
 Thus, the expected number of galaxies with both H$\alpha$ and 
 [OIII] 5007$\mbox{\AA}$ emission lines in this survey is $\sim11$.

 The actual number of galaxies in our sample with possible [OIII]
 and H$\alpha$ emission is eight or nine of the ten double-line
 objects. The line identification for one object is uncertain
 [J033310.66-275221.4a], and the other one [J121900.59+472830.9c]
 is likely a zero-order contaminant. The number is in good agreement
 with the expected $\sim11$ H$\alpha$/[OIII] emitters. 
 Figure \ref{fig:comment} shows one-dimensional spectra of ten objects 
 with plausible double-line features.

\textit{J014107.80-652849.1a} -- 
 This object shows two emission lines at 1.73$\mu$m and 1.32$\mu$m
 although the line at 1.32$\mu$m is only marginally detected.
 These lines are likely redshifted H$\alpha$ and [OIII] at $z=1.63$. 
 We could not derive a reliable flux for [OIII] line. According
 to the equivalent width and the luminosity of H$\alpha$ line, 
 this object is not considered as a Seyfert 1 galaxy (see Section 3.1). 
 The extended morphology (SExtractor CLASS\_STAR $<0.8$) also 
 disfavors the possibility of this object being an AGN. 

\textit{J015240.44+005000.2a} -- 
 This object shows two emission lines at 1.52$\mu$m and 1.16$\mu$m 
 and both lines are significant. The lines are likely 
 redshifted H$\alpha$ and [OIII] at $z=1.31$. The large H$\alpha$
 luminosity and the equivalent width, as well as the point-like morphology
 suggest that this object may be an AGN. The broad and asymmetric shape
 of the line at 1.16$\mu$m could be a blend of [OIII] 5007$\mbox{\AA}$ and
 H$\beta$.

\textit{J033310.66-275221.4a} --
 This object shows two emission lines at 1.32$\mu$m and 1.76$\mu$m, but 
 the identification of these two lines is difficult because the wavelength
 ratio is uncertain. The measured flux ratio 
 between the two lines $EW_{1.76\mu m}$/$EW_{1.32\mu m}$ is less than 1, 
 so H$\alpha$ would be weaker than [OIII]. On the other hand, the two 
 lines may be explained more easily by [OII] and [OIII] lines redshifted 
 to $z=2.52$. The morphology of this object is clearly extended, and the 
 relatively smaller object size (radius of $\sim0.5\arcsec$)
 compared to other emission-line galaxies also suggests the possibility of
 this object at $z>2$. The optical photometry of this object is available 
 in MUSYC data (Gawiser et al. 2006), although the quality of SED fitting
 is low and the derivation of photometric redshift is difficult for this
 object. In Table \ref{tab:objects}, we identify the line at
 1.76$\mu$m as the H$\alpha$ redshifted to $z=1.68$. However, this object 
 is an example of possible mis-identification in construction of emission-line
 galaxy sample, thus we include it in the uncertainties of the luminosity 
 function in Section 4.

\textit{J033310.66-275221.4b} --
 This object shows two emission lines at 1.83$\mu$m and 1.39$\mu$m
 although the line at 1.39$\mu$m is only marginally detected. The lines are
 likely redshifted H$\alpha$ and [OIII] at $z=1.788$. Judging from the clearly
 extended morphology and the combination of EW and luminosity of the H$\alpha$
 line, we conclude that this object is a star-forming galaxy rather
 than AGN. 

\textit{J121900.59+472830.9c} --
 This object has a very bright broad line at 1.77$\mu$m, and a
 weak line at 1.31$\mu$m. These may be redshifted H$\alpha$ and
 H$\beta$ at $z=1.7$. However the line at 1.77$\mu$m is  
 strong (corresponding to a luminosity of 
 $L (\mbox{H}\alpha) = 1.8\times10^{43} \mbox{erg} \mbox{s}^{-1}$, 
 roughly $\sim5L^*$) and broad, so it might be from a zero-order 
 image of an adjacent galaxy. The location of the potential counterpart
 for this possible zero-order image is out of the field of view of our direct
 image, so we cannot confirm clearly whether this is
 a zero-order image or a true emission line. 
 On the other hand, if this line is indeed H$\alpha$, 
 we measure a Balmer decrement of H$\alpha$/H$\beta$ = 9.1, implying 
 $A_{V} = 2.09$.  

\textit{J122512.77+333425.1a} --
 We see two lines at 1.57$\mu$m and 1.19$\mu$m, which we identify as  
 redshifted H$\alpha$ and [OIII] at $z=1.39$. The measured flux ratio
 between the two lines is $f_{1.57\mu m}$/$f_{1.19\mu m}\sim1.06$, 
 which falls within the possible range of H$\alpha$/[OIII] values for 
 star-forming galaxies. 
 Moreover the equivalent width and the luminosity of H$\alpha$ line remove
 the possibility of this object being a Seyfert 1 galaxy. The object shows 
 relatively symmetric but not compact morphology in the direct (F160W) image, 
 thus providing support that this object is a star-forming galaxy rather than
 AGN. 

\textit{J123356.44+091758.4a} --
 This object shows two emission lines at 1.84$\mu$m and 1.38$\mu$m
 although the line at 1.38$\mu$m is very weakly detected. The lines are likely
 redshifted H$\alpha$ and [OIII] at $z=1.8$. The large size of this 
 object, large H$\alpha$ EW and large H$\alpha$ luminosity suggest this
 object is a large galaxy being powered by AGN.

\textit{J125424.89+270147.2a} --
 This object shows two emission lines at 1.88$\mu$m and 1.43$\mu$m.
 The lines are likely redshifted H$\alpha$ and [OIII] at $z=1.861$.
 Judging from the clearly extended morphology and the combination of EW  and 
 luminosity of H$\alpha$ lines, we conclude that this object is a
 star-forming galaxy rather than AGN.

\textit{J161349.94+655049.1a} --
 In addition to the significant line at 1.9$\mu$m, the spectrum of
 this object shows a weak line at 1.43$\mu$m. These lines are likely redshifted 
 H$\alpha$ and [OIII] at $z=1.9$, although the line at 1.43$\mu$m is only
 marginally detected. The line at 1.9$\mu$m has relatively large equivalent
 width (473$\mbox{\AA}$). The spatial scale of the object
 is relatively large and shows a sign of substructure, so we
 classify it as a bright star-forming galaxy.

\textit{J213717.00+125218.5a} --
 This object shows two lines at 1.68$\mu$m and 1.28$\mu$m, which
 are probably redshifted H$\alpha$ and [OIII] at $z=1.56$.
 The flux ratio between the two lines is 
 $f_{1.68\mu m}$/$f_{1.28\mu m}\sim1.55$, although the [OIII] 
 line is weak. 
 The morphology of this object in the F160W image is not classified
 as a point source (with CLASS\_STAR less than 0.9 in SExtractor
 output), although it does not show any clear signs of interaction.

\section{H$\alpha$ LUMINOSITY FUNCTION}

 We use the 1/$V_{\mbox{\tiny{max}}}$ method (Schmidt 1968) in order 
 to construct the H$\alpha$ luminosity function.
 Although the redshift distribution of the galaxies is not homogeneous
 due to the sharp cutoff at low/high redshift end, we find the effect
 is negligible in the calculation of $V_{\mbox{\tiny{max}}}$.
 We begin by calculating the maximum comoving volume $V_{\mbox{\tiny{max}}}$ 
 over which each galaxy could lie and be detected, including corrections 
 for all of our sample selection biases: redshift, 
 flux, and the location in the image. The equation used is as follows 
 (see eqn [1] of Yan et al. 1999) :

\begin{center}
 \begin{equation}
  V_{\mbox{max}}=\Omega (f_{\mbox{H}\alpha})\times \int_{z1}^{z2}
       C(f_{\mbox{H}\alpha}, f_{\mbox{lim}}) R(\lambda_{\mbox{H}\alpha}) 
           \frac{dV}{dz} dz
 \end{equation}
\end{center}

 In the above equation, the differential comoving volume $dV/dz$
 is integrated over the redshift range of [z1, z2].  
 The integration range is defined as
 $z1 = \mbox{max} (z_l, z_{\mbox{min}})$,
 $z2 = \mbox{min} (z_h, z_{\mbox{max}})$ while $z_l$ and $z_h$
 indicate the lower/higher redshift ends limited by the spectral
 cutoff of the G141 response curve ($z_l=0.67$ and $z_h=1.9$).
 The quantities
 $z_{\mbox{min}}$ and $z_{\mbox{max}}$
 are the minimum/maximum redshift the object can be detected.
 While max($z_l, z_{\mbox{min}}$) is $z_l$ in general,
 the maximum redshift the object can be detected is determined 
 by the measured object flux, according to the equation
 $D_L(z_{\mbox{max}}) = D_L(z)
     [f_{\mbox{H}\alpha}/f_{\mbox{lim}}]^{1/2}$ where
 $D_L$ is the luminosity distance at redshift $z$,
 $f_{\mbox{H}\alpha}$ is the measured flux of the identified 
 H$\alpha$ line, and $f_{\mbox{lim}}$ is the 
 limiting line flux of the image (see Section 2.3.).
 The spectral cutoff $z_l$ and $z_h$ vary according to the 
 location of the object in NICMOS grism field of view. Covering
 the whole wavelength range of 1.1--1.9$\mu$m is only possible 
 when the object lies at central portion of the image. Therefore,
 we used different $z_l$ and $z_h$ according to the object location
 in the grism field of view.

\begin{figure*}
  \epsscale{1.1}\plottwo{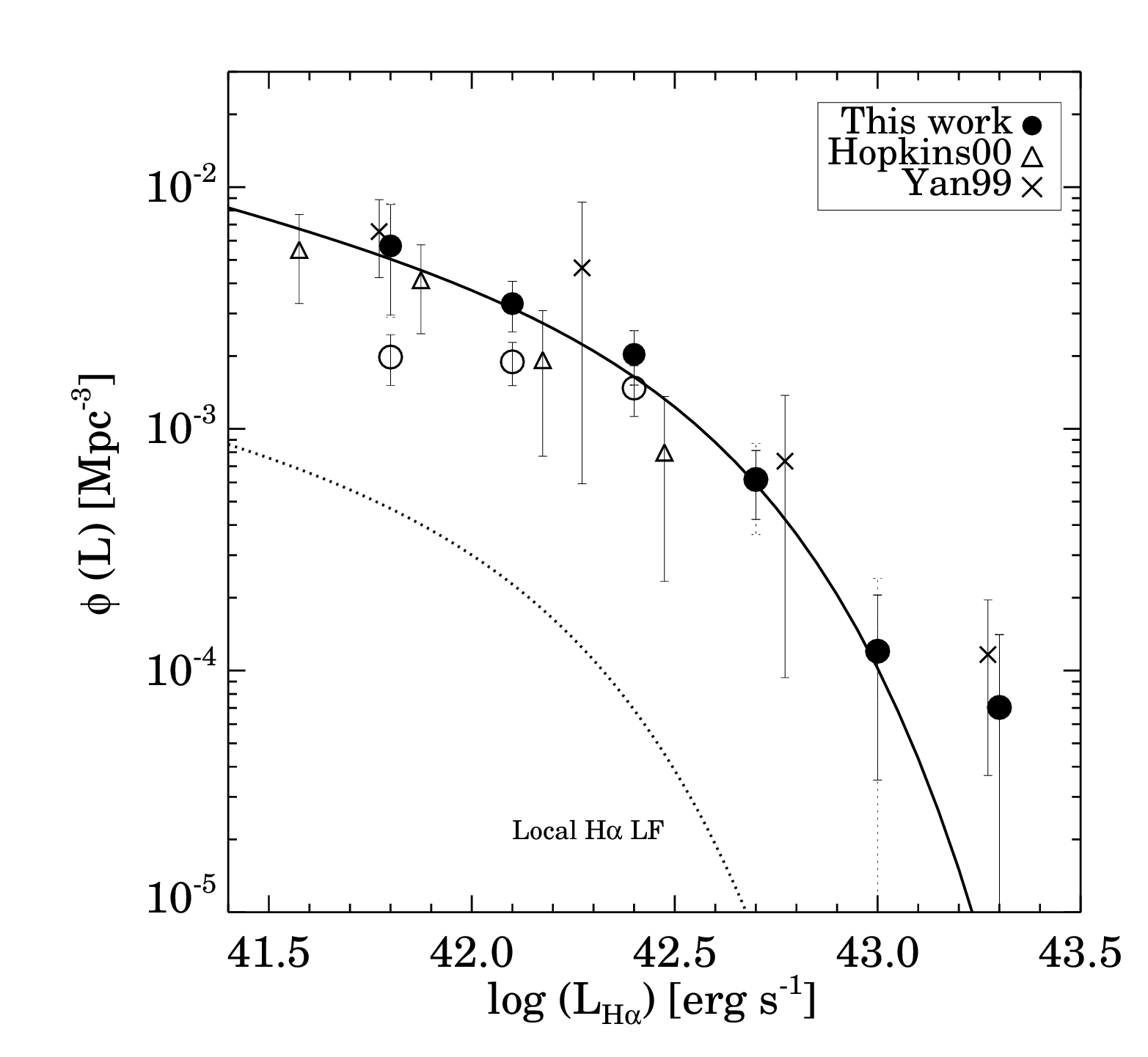}{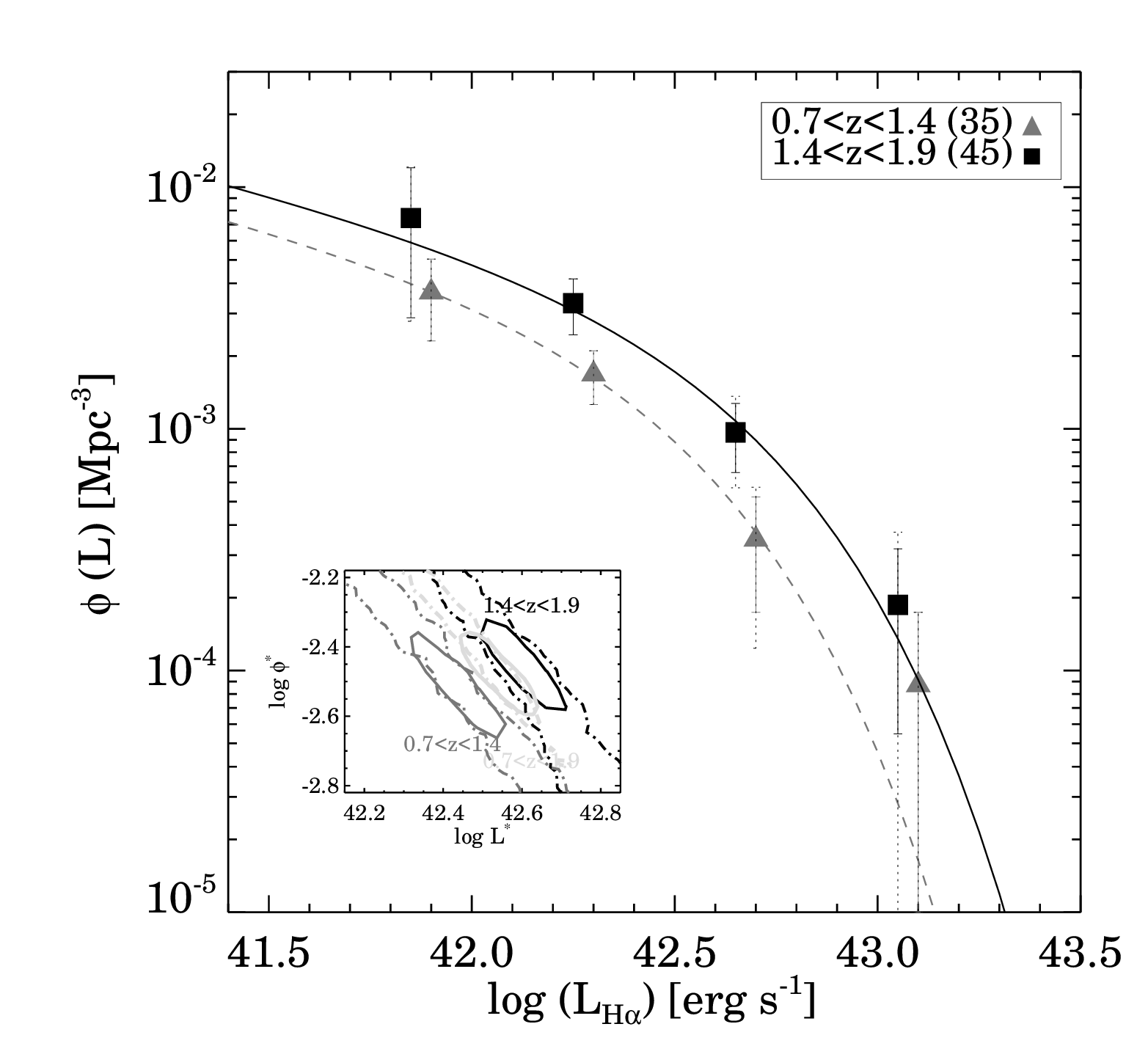}
  \caption{ \label{fig:LFall}
  (\textit{Left}): H$\alpha$ luminosity function over $0.7<z<1.9$ derived
  from our study of emission-line galaxies in NICMOS grism survey.
  The observed values are plotted as \textit{open} circles,
  and the values after the incompleteness correction are
  plotted as \textit{filled} circles. Error bars indicate Poisson
  (\textit{solid}) and additional(\textit{dotted}; uncertainties from
  line mis-identification, contamination) errors. 
  Compared are Yan et al.(1999), Hopkins et al.(2000) points
  also from  NICMOS grism studies
  (\textit{cross/triangle} respectively).
  The \textit{solid} line is the best-fit Schechter
  luminosity function to our derived luminosity function points,
  while the \textit{dotted} line indicates the H$\alpha$ luminosity
  function of local galaxies (Gallego et al. 1995).
  (\textit{Right}): H$\alpha$ luminosity function derived at two
  different redshift range ($0.7<z<1.4$, $1.4<z<1.9$).
  Overplotted \textit{solid} and \textit{dashed} lines are 
  the best-fit Schechter luminosity function with the faint-end
  slope $\alpha$ fixed to $-1.39$. In the inset plot, the best-fit 
  Schechter LF parameters for the two redshift-bin sub-samples are shown 
  in addition to the Schechter parameters for $0.7<z<1.9$ samples. 
  The \textit{solid} contours indicate $1\sigma$ uncertainties in $L^*$ and
  $\phi^*$ when faint-end slope $\alpha$ is fixed, while \textit{dot-dashed}
  contours indicate $1\sigma$ uncertainties when $\alpha$ is a free
  parameter. 
   }
\end{figure*}

 The observed luminosity function needs to be corrected for sample
 selection biases, including redshift limits and the incompleteness.
 $R(\lambda)$ is the G141 inverse sensitivity curve,
 which corrects for the effect of the sharp spectral cutoff at the low-
 and high-redshift ends.
 Since we are using images with different depths, different 
 incompleteness corrections must be applied for individual objects.
 $C (f_{\mbox{H}\alpha}, f_{\mbox{lim}}) $ is a factor
 for incompleteness correction assigned to each object, as a function 
 of the observed H$\alpha$ line flux and the flux limit of the image. 
 To estimate this correction factor, 
 we first select several
 high S/N emission-line galaxies with different equivalent widths and 
 wavelengths. After cutting out two-dimensional spectral templates of 
 the selected emission-line galaxies, we dim the images by various factors
 to generate artificial emission-line galaxies with various line fluxes. 
 The dimmed two-dimensional spectra are added to original NICMOS grism
 images at random locations, then we perform the one-dimensional 
 spectral extraction, applying the same method that we described in
 Section 2.2.
 We repeat these steps for images with different $f_{\mbox{lim}}$
 to measure $C (f_{\mbox{H}\alpha}, f_{\mbox{lim}})$ as a function of
 galaxy location.
 We plot both uncorrected and corrected luminosity function 
 in Figure \ref{fig:LFall}a in order to show the significance of the 
 incompleteness correction. 
 Finally, $\Omega$ is the solid angle covered in this survey. 
 $\Omega$ is also a function of $f_{\mbox{H}\alpha}$ since our 
 objects are gathered from multiple images with widely differing line detection
 depths. 

 The source density in a specific luminosity bin of width 
 $\Delta (\mbox{log} L)$ centered on the luminosity $\mbox{log} L_i$
 is the sum of (1/$V_{\mbox{max}}$) of all sources
 within the luminosity bin (i.e.,
 $\mbox{log} L_{i+1} - \mbox{log} L_{i} = 2 \times \Delta (\mbox{log} L)$).
 The variances are computed by summing the squares of the inverse
 volumes, thus the luminosity values and the error-bar in each bin 
 is evaluated using the following equations.

\begin{center}

 \begin{equation}
  \phi(\mbox{log} L_i) = \frac{1}{\Delta (\mbox{log} L)}
     \sum_{| \mbox{log} L - \mbox{log} L_i | < \Delta (\mbox{log} L) } 
      \frac{1}{V_{\mbox{max}}} 
 \end{equation}

 \begin{equation}
  \sigma_{\phi} = \frac{1}{\Delta \mbox{log} L}
     \sqrt {\sum_{| \mbox{log} L - \mbox{log} L_i | < \Delta (\mbox{log} L) } 
      \left(\frac{1}{V_{\mbox{max}}}\right)^2 }
 \end{equation}

\end{center}

 After calculating the size of the error-bars using the equation,
 we added additional uncertainties that might result from AGN
 contamination and line mis-identification (please refer to Section 3.1
 for details). Note that the uncertainties in the luminosity function and
 the derived SFR density due to the large scale structure are less than 
 $\sim2$\% (Trenti \& Stiavelli 2008), given the large number of 
 independent fields used in our survey.
 The derived luminosity functions from the whole sample ($0.7<z<1.9$)
 is presented in
 Figure \ref{fig:LFall}a and Table \ref{tab:LF_val}, while the 
 errors indicated already include additional uncertainties from 
 AGN contamination and line mis-identification.
 The points from the previous NICMOS grism selected H$\alpha$ emitters
 (Yan et al. 1999; Hopkins et al. 2000) are overplotted for 
 comparison, after accounting for the cosmology differences.
 Over the luminosity range of 
 $41.6 < \mbox{log} L (\mbox{H}\alpha) [\mbox{erg s}^{-1}] <43.5$, our LF
 is consistent with those of the previous studies. The solid line in Figure
 \ref{fig:LFall}a is the best-fit Schechter LF to 
 our data points, yet the line still fits points from other 
 studies as well within the error bars. The plot shows a clear
 evolution of the H$\alpha$ LF from $z=1.4$ to the local H$\alpha$ LF
 (dotted line), a result already well-known from the luminosity 
 evolution in other wavelengths like the UV (e.g., Arnouts et al. 2005).
 
 The H$\alpha$ luminosity function we have derived is for galaxies
 with EW (rest-frame) $>40\mbox{\AA}$, as our survey is unable to   
 confidently detect lines with lower equivalent widths.  However, we note
 that our observed range of EWs is comparable to that of $z\sim2$
 star-forming galaxies observed in Erb et al.(2006).  This suggests that
 there may be substantial overlap between our sample and the one reported
 by Erb et al.(2006), although our sample likely contains some dust obscured
 galaxies that would be missed by the UV selection of Erb et al.(2006). 

 The best-fit Schechter function parameters (Table \ref{tab:lfparam})
 are derived using the \texttt{MPFIT}
 package\footnote{http://cow.physics.wisc.edu/\ensuremath{\sim}craigm/idl/fitting.html},
 which provides a robust non-linear least square curve fitting (e.g., Ly et al. 2007). 
 The errors in Table \ref{tab:lfparam} correspond to $1\sigma$
 uncertainty for each Schechter function parameter, and are derived
 using a Monte Carlo simulation. We generated a large number ($\sim10000$)
 of Monte Carlo realizations of our LF, with different [log$L$, log$\phi$]
 sets perturbed according to the uncertainties. 
 Then we repeat the fit to find the best-fit parameters for each realization
 of the LF. We find the faint-end slope is largely unconstrained by our data:
 $\alpha= -1.39\pm0.43$. This is consistent with both the local H$\alpha$ LF
 ($\alpha = -1.35$; Gallego et al. 1995), and the
 deep NICMOS grism H$\alpha$ survey ($\alpha =-1.86\pm0.14$; Hopkins et al. 2000). 
 In addition to the Schechter function fitting with varying $\alpha$, 
 we derived $L^*$ and $\phi^*$ with $\alpha$ being fixed to $-1.39, -1.0$ 
 and $-1.8$ (Table \ref{tab:lfparam}) to cover all the possible range of $\alpha$,
 and to investigate the effect of varying $\alpha$ on the total SFR density derived.

 Since the number of our sample galaxies (80) is more than twice
 of that from the previous studies, we can also test the evolution of 
 H$\alpha$ LF as a function of redshift between 0.7 and 1.9.
 We divided the sample galaxies into two redshift bins 
 ($0.7<z<1.4$, $1.4<z<1.9$), and derived the LFs separately.
 The LFs for two sub-samples at different redshift bin are shown in 
 Figure \ref{fig:LFall}b. 
 The luminosity function in the lower-redshift bin
 ($0.7<z<1.4$; $\langle z \rangle=1.1$) is plotted as triangles,
 while the luminosity function for the higher-redshift bin
 ($1.4<z<1.9$; $\langle z \rangle=1.6$) is plotted as squares. 
 We see that there are more H$\alpha$-luminous galaxies
 in the higher-redshift bin than in the lower-redshift bin, which implies
 that $L^*$, $\phi^*$, or both are larger at higher redshift.

 The LF values and the best-fit Schechter parameters for these two 
 redshift bins are also listed in Table \ref{tab:LF_val} and \ref{tab:lfparam}.
 We also show the derivation of the Schechter parameters for the faint-end
 slope fixed at $\alpha=-1.39, -1.0, {\rm and} -1.8$ as this quantity is
 more difficult to constrain for these smaller sub-samples. The errors in 
 $\phi^*$ and $L^*$ are derived by the same Monte Carlo method described above.
 For the cases where $\alpha$ is fixed, the uncertainties are artificially
 decreased, so we adopt the larger uncertainties derived when $\alpha$ is free. 
 In the inset plot of Figure \ref{fig:LFall}b, we illustrated how $\phi^*$
 and $L^*$ evolve from $1.4<z<1.9$ to $0.7<z<1.4$. The contours represent
 the $1\sigma$ uncertainty range for the parameters (for $\alpha$ free and
 fixed, dot-dashed/solid line). The contours for parameters 
 at $0.7<z<1.9$ is also illustrated.

\section{EVOLUTION OF SFR DENSITY at $1<z<2$}  

 From the derived LF parameters, we evaluate the H$\alpha$-inferred
 star formation rate density at $0.7<z<1.4$ (lower-redshift bin), and 
 $1.4<z<1.9$ (higher-redshift bin). The SFR densities are also listed
 in Table \ref{tab:lfparam}.
 The total integrated H$\alpha$ luminosity density
 is calculated as $L_{\mbox{\tiny{tot}}}=\phi^* L^* \Gamma(\alpha+2) $,
 then converted to SFR density using the relation of Kennicutt (1998) :
 $SFR(H\alpha) [\mbox{M}_{\odot} \mbox{yr}^{-1}] = 7.9\times10^{-42} L(H\alpha) [\mbox{erg} \mbox{s}^{-1}]$.
 The relation assumes a Salpeter IMF between 0.1--100 M$_{\odot}$.
 The uncertainties in Table \ref{tab:lfparam} are also derived through
 Monte Carlo method, by making a large realization of [$\phi^*$, $L^*$,
 $\alpha$] sets, calculating $L_{\mbox{\tiny{tot}}}$, and estimating
 $1\sigma$ uncertainty from the distribution of $L_{\mbox{\tiny{tot}}}$.
 As mentioned in previous section, the uncertainties on $L^*$ are 
 underestimated for the cases where $\alpha$ is fixed.

\begin{figure*}
 \epsscale{1.0}\plotone{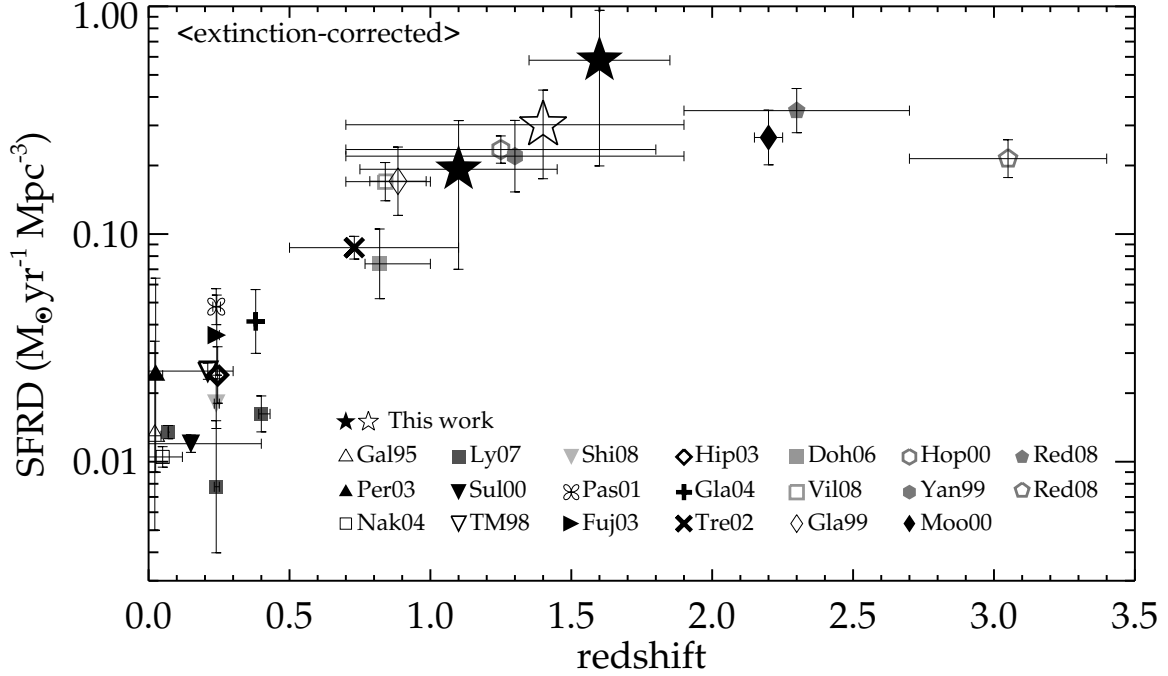}
 \caption{\label{fig:sfrd_evol} The evolution of SFR density as a
 function of redshift.
 Our points are plotted as two \textit{filled}
 stars at $z=1.1$ and 1.6, produced from the integrated luminosity 
 functions of the two sub-samples covering these redshift ranges.
 Also, the SFR density evaluated from the luminosity function
 over the entire range of $0.7<z<1.9$ is plotted at $z=1.4$ (\textit{open}
 star). All other plotted points are based on the
 H$\alpha$-derived SFR density from spectroscopy or narrow-band 
 imaging, (e.g., Gallego et al. 1995; Yan et al. 1999; Hopkins et al. 2000;
 Moorwood et al. 2000; Perez-Gonzalez et al. 2003;
 Sullivan et al. 2000; Tresse \& Maddox 1998; Tresse et al. 2002;
 Pascual et al. 2001; Fujita et al. 2003; Nakamura et al. 2004;
 Hippelein et al. 2003; Glazebrook et al. 1999, 2004; Doherty et al. 2006;
 Ly et al. 2007; Shioya et al. 2008; Villar et al. 2008)
 except for points at $z=2.3$ and 3.05 (Reddy et al. 2008) 
 who used H$\alpha$ luminosity function converted from UV luminosity
 function using a UV-H$\alpha$ relation.
 }
\end{figure*}

 We compare our SFR density estimates with the results of other
 studies in Figure \ref{fig:sfrd_evol}. The SFR points are drawn from
 individual references (Gallego et al. 1995; Yan et al. 1999;
 Hopkins et al. 2000; Moorwood et al. 2000; Perez-Gonzalez et al. 2003;
 Sullivan et al. 2000; Tresse \& Maddox 1998; Tresse et al. 2002;
 Pascual et al. 2001; Fujita et al. 2003; Nakamura et al. 2004;
 Hippelein et al. 2003; 
 Glazebrook et al. 1999, 2004; Doherty et al. 2006;
 Ly et al. 2007; Shioya et al. 2008; Villar et al. 2008;
 Reddy et al. 2008; see compilation of Hopkins 2004), and were corrected 
 to fit different cosmology. For extinction, we used the values given by
 the authors. The amount of the applied extinction 
 correction is different for different studies, from $A(H\alpha)\sim0.3$ mag 
 to $A(H\alpha)\sim1.2$ mag. This difference up to 1 magnitude produce the
 uncertainty in the SFR density points of a factor of $\sim2$. 
 If the authors do not provide
 information about the extinction, we apply the average extinction of 
 $\langle A_V \rangle=1$ mag, i.e., $\langle A(H\alpha) \rangle\sim0.85$ mag
 (e.g., Kennicutt 1992), the factor widely adopted in previous studies (e.g.,
 Hopkins 2004; Doherty et al. 2006). The extinction correction applied
 to our SFR density points (stars in Figure \ref{fig:sfrd_evol}) is also
 $\langle A(H\alpha) \rangle\sim0.85$ mag.

 Our estimate of the volume-averaged SFR density at $0.7<z<1.9$ is
 consistent with that of previous NICMOS grism studies over the same
 redshift range (points at $z=1.3$ from Yan et al. 1999;
 $z=1.25$ from Hopkins et al. 2000). Note that the integration ranges for
 H$\alpha$ luminosity density are different from study to study --
 our study accepts $L$ in the range $[0,\infty]$ while Hopkins et al.(2000)
 have integrated the LF over the range of $10^{37} < L [\mbox{erg}] < 10^{47}$.
 Since the derived faint-end slope in our study is relatively flat ($1.39\pm0.43$),
 this difference of integration range makes little difference in the final
 SFR density value. If we restrict the integration range to 
 $10^{37} < L [\mbox{erg}] < 10^{47}$ as in Hopkins et al.(2000), our result
 is decreased by only $\sim0.1$\%. 

 Our points clearly place the peak epoch of the relatively ``unobscured'' 
 star formation at $z=1$--2. Moreover, by examining the two points at $z=1.1$ and
 $z=1.6$, we can conclude that the peak of star formation must have 
 occurred at redshifts higher than $z=1.5$. This result is fairly robust,
 as it uses two sets of galaxies all identified
 with the same selection method.

\section{SUMMARY}

 We have designed and executed a NICMOS grism survey, exploring the rest-frame 
 optical universe at $0.7<z<1.9$. Through
 this program, we have identified a unique sample of emission-line galaxies
 over a significant cosmic volume, which enable us to study the
 relatively bright part of the H$\alpha$ luminosity function
 at $0.7<z<1.9$. 

 Using Cycle 12 and Cycle 13 data, we probe $\sim104$ arcmin$^2$
 area, at 139 different locations throughout the sky. This corresponds
 to an effective comoving volume of $\sim3.14\times10^5 \mbox{Mpc}^3$,
 almost two times larger than that of our previous observations
 (McCarthy et al. 1999). 
 We identified 80 probable emission-line galaxies, down to 
 $L (\mbox{H}\alpha)\sim 4.4\times10^{41} \mbox{erg} \mbox{s}^{-1}$.
 Most of the emission lines
 are thought to be redshifted H$\alpha$, and their $H$-band
 magnitude distribution suggests that the identified emission-line galaxies
 are relatively bright ($M\sim M^*$) star-forming galaxies at 
 $0.7<z<1.9$.
 
 We construct the H$\alpha$ luminosity function from the line fluxes and 
 the redshifts of these galaxies.
 From the integration of the luminosity function, the luminosity density
 and the star formation rate density are derived. 
 We divide our sample into two redshift bins, one at $0.7<z<1.4$
 and the other at $1.4<z<1.9$. The volume-averaged star formation rate
 densities at these two different redshift range are evaluated 
 to be $0.088\pm0.056 \mbox{M}_{\odot} \mbox{yr}^{-1} \mbox{Mpc}^{-3}$ and
 $0.265\pm0.174 \mbox{M}_{\odot} \mbox{yr}^{-1} \mbox{Mpc}^{-3}$ respectively. 
 The results are consistent with other H$\alpha$-derived SFR densities
 at similar redshifts. Using our unique sample, all selected by the same method,
 we find that the cosmic star formation history probed by these H$\alpha$
 measurements places the peak of star formation at $z>1.5$,
 with a decrease in global SFR density from $z=1.6$ to $z=1.1$.  
 Although there remain uncertainties
 in the relative extinction, our study places firm constraints
 on the cosmic star formation history.

\clearpage

\begin{deluxetable}{lccccccccrr}
 \tabletypesize{\scriptsize}
 \tablewidth{0pt}
 \setlength{\tabcolsep}{0.01in}
 \tablecaption{\label{tab:objects} Emission-line objects from
 the NICMOS grism survey  }
 \tablehead{
  \colhead{Field} & \colhead{Object\tablenotemark{a}} &
  \colhead{$\alpha$ (J2000)} & \colhead{$\delta$ (J2000)} &
  \colhead{redshift$_{H\alpha}$   } &
  \colhead{Flux$_{H\alpha}$\tablenotemark{b}} & \colhead{$W_{obs}$} &
  \colhead{F110W\tablenotemark{c}} & \colhead{F160W\tablenotemark{c}} &
  \colhead{logL$_{\mbox{H}\alpha}$\tablenotemark{d}} & \colhead{SFR\tablenotemark{d}}
  \\
  \colhead{} & \colhead{} &
  \colhead{} & \colhead{} &
  \colhead{} &
  \colhead{($10^{-16}$ erg s$^{-1}$ cm$^{-2}$)} & \colhead{($\mbox{\AA}$)} &
  \colhead{(mag)} & \colhead{(mag)} &
  \colhead{(erg s$^{-1}$)} & \colhead{($\mbox{M}_{\odot} \mbox{yr}^{-1}$ )}
  }
 \startdata
   \tableline
   J002638.02+170117.9 & a & 00:26:38.32 & +17:01:20.7 & 1.206 &  2.653 & 320 & 23.13 & 22.52 & 42.24 & 13.8 \\
   \nodata             & b & 00:26:38.35 & +17:01:24.8 & 1.300 &  1.299 & 230 & 23.48 & 22.61 & 41.97 &  7.4 \\
   J013433.61+311818.2 & a & 01:34:34.38 & +31:18:05.3 & 1.283 &  3.372 & 283 & 22.20 & 22.20 & 42.38 & 19.0 \\
   \nodata             & b & 01:34:34.85 & +31:17:56.0 & 1.657 &  5.546 & 281 & 24.47 & 23.26 & 42.73 & 42.8 \\
   J013510.41+313024.1 & a & 01:35:10.50 & +31:29:31.3 & 1.441 &  1.545 & 350 & -- & 23.42 & 42.10 & 10.0 \\
   J013655.80+153720.4 & a & 01:36:54.82 & +15:37:27.8 & 1.083 &  2.827 & 139 & 22.72 & 22.21 & 42.21 & 12.9 \\
   \nodata             & b & 01:36:55.37 & +15:37:51.5 & 0.773 &  3.645 & 334 & 21.93 & 21.57 & 42.14 & 11.0 \\
   J013916.84-002913.1 & a & 01:39:14.18 & -00:28:38.9 & 0.888 &  2.870 & 185 & -- & 22.29 & 42.11 & 10.2 \\
   \nodata             & b & 01:39:15.21 & -00:28:23.6 & 1.407 &  4.291 & 322 & -- & 21.24 & 42.53 & 27.1 \\
   \nodata             & c & 01:39:16.38 & -00:28:33.6 & 1.651 &  2.453 & 244 & -- & 21.68 & 42.38 & 18.8 \\
   J014107.80-652849.1 & a & 01:41:06.40 & -65:28:25.6 & 1.641 &  5.248 & 249 & -- & 22.59 & 42.70 & 40.0 \\
   \nodata             & b & 01:41:08.33 & -65:28:26.1 & 1.047 &  6.941 & 172 & -- & 23.80 & 42.58 & 30.4 \\
   J015108.44-832207.1 & a & 01:51:20.24 & -83:22:00.7 & 1.593 &  2.706 & 254 & 22.67 & 23.39 & 42.40 & 19.9 \\
   J015240.44+005000.2 & a & 01:52:38.65 & +00:50:45.0 & 1.314 &  4.811 & 271 & -- & 21.59 & 42.55 & 27.9 \\ 
   J021011.32-043808.9 & a & 02:10:09.66 & -04:38:14.3 & 1.128 &  2.386 & 148 & -- & 22.88 & 42.16 & 11.4 \\
   J021017.37-043616.3 & a & 02:10:17.23 & -04:36:12.4 & 1.564 &  4.152 & 467 & 23.57 & 22.90 & 42.58 & 29.8 \\
   \nodata             & b & 02:10:17.68 & -04:36:36.5 & 1.043 &  0.947 & 122 & 24.43 & 23.59 & 41.72 &  4.1 \\
   J031951.03-191624.3 & a & 03:19:50.77 & -19:16:46.2 & 0.974 &  4.848 & 217 & 21.87 & 21.36 & 42.38 & 19.4 \\
   J033309.36-275242.1 & a & 03:33:06.60 & -27:52:05.4 & 1.370 &  0.897 & 287 & -- & 24.59 & 41.84 &  5.5 \\
   \nodata             & b & 03:33:07.59 & -27:52:40.9 & 1.590 &  1.779 & 285 & -- & 23.12 & 42.22 & 13.0 \\
   J033310.66-275221.4 & a & 03:33:07.68 & -27:51:47.1 & 1.681 &  0.881 & 135 & -- & 24.63 & 41.94 &  6.9 \\
   \nodata             & b & 03:33:08.27 & -27:51:47.8 & 1.788 &  1.482 & 334 & -- & 24.02 & 42.20 & 12.6 \\
   J034932.44-533706.8 & a & 03:49:34.40 & -53:37:23.1 & 1.465 &  4.250 & 325 & 22.48 & 21.45 & 42.55 & 28.2 \\
   J051917.75-454124.2 & a & 05:19:17.28 & -45:41:30.9 & 1.648 &  0.462 & 136 & 23.20 & 21.96 & 41.65 &  3.5 \\
   \nodata             & b & 05:19:18.36 & -45:40:59.7 & 1.169 &  2.792 & 248 & 22.40 & 21.83 & 42.25 & 14.0 \\
   J054709.71-505535.4 & a\tablenotemark{e} & 05:47:09.51 & -50:55:56.9 & 1.110 &  1.410 & 228 & -- & -- & 41.92 &  6.6 \\
   J081955.60+421755.4 & a & 08:19:55.53 & +42:17:50.5 & 1.122 &  0.958 & 156 & 23.61 & 22.56 & 41.76 &  4.6 \\
   J084830.53+444456.9 & a & 08:48:33.21 & +44:45:27.4 & 1.445 &  2.663 & 284 & -- & 22.56 & 42.34 & 17.4 \\
   \nodata             & b & 08:48:33.66 & +44:45:33.4 & 1.073 &  2.936 & 208 & -- & 22.49 & 42.22 & 13.2 \\
   J084846.69+444336.8 & a & 08:48:47.04 & +44:43:37.1 & 0.986 &  1.591 & 222 & 22.36 & 21.97 & 41.91 &  6.5 \\
   J085828.67-161442.3 & a & 08:58:27.27 & -16:14:30.2 & 1.158 &  1.252 & 180 & 23.46 & 22.45 & 41.90 &  6.2 \\
   \nodata             & b & 08:58:28.93 & -16:14:46.6 & 1.206 &  7.510 & 315 & 22.08 & 21.42 & 42.69 & 39.1 \\
   J091100.06+173832.2 & a & 09:11:00.87 & +17:38:56.5 & 1.333 &  2.795 & 184 & 22.72 & 21.91 & 42.32 & 16.5 \\
   J094841.23+673041.9 & a & 09:48:36.87 & +67:30:46.7 & 1.435 &  2.310 & 331 & -- & 22.11 & 42.28 & 14.9 \\
   \nodata             & b & 09:48:43.91 & +67:30:59.2 & 1.180 &  1.401 & 256 & -- & 24.74 & 41.95 &  7.1 \\
   J100603.06+350241.2 & a & 10:06:01.50 & +35:02:42.9 & 1.620 &  0.914 & 430 & -- & 24.27 & 41.94 &  6.9 \\
   \nodata             & b & 10:06:04.91 & +35:02:45.1 & 1.081 &  4.423 & 109 & -- & 22.84 & 42.41 & 20.1 \\
   J103316.56+231102.6 & a & 10:33:17.25 & +23:11:54.5 & 1.519 &  0.956 & 126 & -- & 23.03 & 41.92 &  6.6 \\
   \nodata             & b & 10:33:19.19 & +23:11:40.0 & 1.372 & 12.496 & 298 & -- & 22.84 & 42.99 & 76.4 \\
   J104703.10+122919.7 & a & 10:47:01.41 & +12:30:06.1 & 0.811 &  4.687 & 265 & -- & 20.82 & 42.28 & 15.0 \\
   \nodata             & b & 10:47:02.36 & +12:29:35.6 & 1.482 &  3.175 & 542 & -- & 21.79 & 42.43 & 21.3 \\
   J104849.15+462852.5 & a & 10:48:48.37 & +46:28:57.0 & 1.632 &  3.292 & 445 & -- & 23.49 & 42.50 & 24.9 \\
   J111942.69+513839.1 & a & 11:19:40.61 & +51:38:55.8 & 1.733 &  1.689 & 219 & 23.09 & 22.00 & 42.24 & 13.8 \\
   J112413.26-170215.6 & a & 11:24:13.50 & -17:02:14.1 & 1.744 &  0.719 & 185 & 23.94 & 23.47 & 41.87 &  5.9 \\
   \nodata             & b & 11:24:13.70 & -17:02:33.2 & 1.349 &  2.517 & 443 & 23.53 & 22.34 & 42.28 & 15.1 \\
   J112414.50-170137.3 & a & 11:24:15.58 & -17:01:45.9 & 1.667 &  1.041 & 185 & 23.79 & 23.02 & 42.01 &  8.1 \\
   J112846.94+641459.7 & a & 11:28:44.19 & +64:15:06.3 & 1.005 &  1.467 & 137 & 23.62 & 23.75 & 41.89 &  6.1 \\
   \nodata             & b & 11:28:48.60 & +64:15:00.5 & 1.559 &  2.655 & 177 & 22.58 & 23.33 & 42.38 & 19.0 \\
   J120513.71-073124.0 & a & 12:05:13.33 & -07:30:54.9 & 1.550 &  0.722 & 190 & -- & 25.18 & 41.81 &  5.1 \\
   J121900.59+472830.9 & a\tablenotemark{e} & 12:18:58.73 & +47:28:16.6 & 1.679 &  1.751 & 236 & -- & -- & 42.24 & 13.7 \\
   \nodata             & b & 12:18:59.24 & +47:28:10.9 & 1.462 &  4.513 & 115 & 22.92 & 22.16 & 42.58 & 29.8 \\
   \nodata             & c & 12:18:59.48 & +47:28:10.8 & 0.996 &  3.779 & 200 & 21.63 & 21.06 & 42.29 & 15.5 \\
   J121936.36+471950.6 & a & 12:19:35.78 & +47:19:35.8 & 1.186 &  1.391 & 220 & -- & 24.56 & 41.95 &  7.1 \\
   J122226.64+042914.2 & a & 12:22:25.25 & +04:28:35.9 & 1.601 &  0.556 & 106 & -- & 22.88 & 41.72 &  4.1 \\
   J122246.79+155704.0 & a & 12:22:47.16 & +15:56:45.2 & 1.697 &  0.853 & 126 & 22.94 & 22.09 & 41.93 &  6.8 \\
   J122512.77+333425.1 & a & 12:25:11.12 & +33:34:31.7 & 1.392 &  1.950 & 148 & 22.68 & 21.72 & 42.19 & 12.1 \\
   \nodata             & b & 12:25:13.26 & +33:34:14.3 & 1.479 &  1.402 & 209 & 23.89 & 22.55 & 42.08 &  9.4 \\
   J122908.41+015420.1 & a & 12:29:08.56 & +01:54:08.5 & 1.752 &  1.868 & 171 & -- & 20.96 & 42.29 & 15.4 \\
   J123041.86+121509.1 & a & 12:30:42.28 & +12:15:26.9 & 1.475 &  2.437 & 188 & -- & 22.40 & 42.31 & 16.3 \\
   J123356.44+091758.4 & a & 12:33:56.24 & +09:17:54.0 & 1.800 & 14.161 & 371 & -- & 20.75 & 43.18 &120.9 \\
   J124339.45-341843.1 & a & 12:43:40.88 & -34:18:20.6 & 1.050 &  3.115 & 146 & 22.58 & 21.87 & 42.24 & 13.7 \\
   J125424.89+270147.2 & a & 12:54:24.63 & +27:02:05.2 & 1.861 &  2.837 & 211 & -- & 22.51 & 42.50 & 25.2 \\
   J132745.26-311537.8 & a & 13:27:44.50 & -31:16:07.5 & 1.240 &  2.566 & 251 & 23.56 & 23.13 & 42.24 & 13.8 \\
   J132820.08-313744.8 & a & 13:28:19.70 & -31:37:42.2 & 1.280 &  1.474 & 274 & 22.68 & 21.84 & 42.02 &  8.3 \\
   J135835.94+623046.5 & a & 13:58:38.24 & +62:30:55.8 & 1.476 &  0.638 & 203 & 23.44 & 22.88 & 41.73 &  4.3 \\
   J140214.22-113634.5 & a & 14:02:13.67 & -11:36:53.1 & 1.514 &  0.711 & 215 & 23.09 & 22.85 & 41.79 &  4.9 \\
   J141833.44+250745.0 & a & 14:18:33.84 & +25:07:45.1 & 1.419 &  1.629 & 215 & 24.48 & 24.39 & 42.12 & 10.4 \\
   J161349.94+655049.1 & a & 16:13:52.19 & +65:50:51.4 & 1.904 &  6.688 & 473 & -- & 23.66 & 42.89 & 61.2 \\
   \nodata             & b\tablenotemark{e} & 16:13:47.61 & +65:50:35.0 & 1.748 &  0.692 & 195 & -- & -- & 41.86 &  5.7 \\
   J175907.13+664454.3 & a & 17:59:03.06 & +66:45:13.0 & 1.331 &  1.620 & 178 & 24.56 & 24.72 & 42.08 &  9.5 \\  
   \nodata             & b & 17:59:09.47 & +66:44:48.9 & 1.614 &  1.618 & 190 & 25.12 & 24.13 & 42.18 & 12.1 \\
   \nodata             & c & 17:59:11.37 & +66:44:41.9 & 1.446 &  0.640 & 131 & 24.11 & 23.67 & 41.72 &  4.2 \\
   J213717.00+125218.5 & a & 21:37:18.38 & +12:51:34.8 & 1.561 &  6.620 & 226 & -- & 20.85 & 42.78 & 47.4 \\
   J220239.22+185112.4 & a & 22:02:38.87 & +18:51:37.9 & 1.406 &  2.759 & 312 & -- & 23.40 & 42.34 & 17.4 \\
   \nodata             & b & 22:02:39.09 & +18:51:29.3 & 1.505 &  4.187 & 484 & 23.72 & 23.13 & 42.56 & 28.7 \\
   \nodata             & c & 22:02:40.03 & +18:51:24.5 & 1.663 &  1.678 & 242 & 24.62 & 23.60 & 42.22 & 13.0 \\
   J222640.90-722919.8 & a & 22:26:41.14 & -72:29:14.3 & 1.692 &  1.477 & 191 & -- & 20.95 & 42.17 & 11.7 \\
   J225916.39-345408.5 & a & 22:59:16.51 & -34:54:11.5 & 1.507 &  2.931 & 254 & -- & 22.39 & 42.41 & 20.1 \\
   J230342.56+085617.2 & a & 23:03:42.20 & +08:56:42.9 & 1.355 &  5.527 & 156 & 22.46 & 21.59 & 42.62 & 33.3 \\
   \nodata             & b & 23:03:43.68 & +08:56:07.4 & 1.164 &  0.799 & 581 & 23.33 & 23.00 & 41.70 &  4.0
 \enddata
 \tablenotetext{a}{We assign the suffix a, b, or c to identify 
 different objects in one field.}
 \tablenotetext{b}{Flux$_{H\alpha}$ is the emission line flux. Since the 
 [NII] and H$\alpha$ lines are not resolved in the resolution of NICMOS grism, 
 Flux$_{H\alpha}$ represents $f(H\alpha+[\mbox{NII}])$. }
 \tablenotetext{c}{F110W/F160W magnitudes are in AB magnitudes, 
 MAG\_AUTO from \textit{SExtractor} (total magnitudes of the galaxies).
 The conversion between AB and Vega magnitudes are
 $\mbox{F110W}_{Vega}=\mbox{F110W}_{AB}-0.73$;
 $\mbox{F160W}_{Vega}=\mbox{F160W}_{AB}-1.31$
 (derived using NICMOS zeropoints in Vega magnitude system at
  http://www.stsci.edu/hst/nicmos/performance/photometry/postncs\_keywords.html).
 }
 \tablenotetext{d}{H$\alpha$ luminosity and the derived star
 formation rate are corrected for possible [NII] contamination
 in the measured H$\alpha$ line flux:
 $f(H\alpha)=0.71\times f(H\alpha + [\mbox{NII}]) $
 }
 \tablenotetext{e}{The objects that do not have either F110W/F160W photometry
  are objects lying near the image edge. Due to 
  the locations, we could not find these objects in the direct image despite
  of clear emission lines (not thought to be zero order) for these galaxies. 
  All galaxies without F110W photometry do not have the corresponding
  F110W images, i.e., there is no F110W-dropouts or galaxies with very red 
  (F110W$-$F160W) colors. }
\end{deluxetable}

\begin{deluxetable}{lcr clcr clcr}
 \tablewidth{0pt}
 \tablecaption{\label{tab:LF_val}
  The derived H$\alpha$ luminosity function in grism survey}
 \tablehead{
   \multicolumn{3}{c}{ $z\sim$ 0.7--1.9} &
   \colhead{} & \multicolumn{3}{c}{ $z\sim$ 0.7--1.4} &
   \colhead{} & \multicolumn{3}{c}{ $z\sim$ 1.4--1.9} \\
 \cline{1-3}  \cline{5-7}  \cline{9-11} \\
   \colhead{log$L$} &  \colhead{$\phi$ (10$^{-3} \mbox{Mpc}^{-3}$)} & 
   \colhead{$N_{gal}$} &
   \colhead{} &
   \colhead{log$L$} &  \colhead{$\phi$ (10$^{-3} \mbox{Mpc}^{-3}$)} & 
   \colhead{$N_{gal}$} &
   \colhead{} &
   \colhead{log$L$} &  \colhead{$\phi$ (10$^{-3} \mbox{Mpc}^{-3}$)} & 
   \colhead{$N_{gal}$} }
 \startdata
    41.8 & $5.7115\pm2.8134$ & 20 &  & 41.9 & $3.6746\pm1.3633$ & 13 &  & 41.85 & $7.4484\pm4.6620$ & 13 \\
    42.1 & $3.2962\pm0.7774$ & 26 &  & 42.3 & $1.6812\pm0.4203$ & 17 &  & 42.25 & $3.3088\pm0.8610$ & 20 \\
    42.4 & $2.0319\pm0.5149$ & 21 &  & 42.7 & $0.3485\pm0.2248$ &  4 &  & 42.65 & $0.9663\pm0.3960$ & 10 \\
    42.7 & $0.6172\pm0.2522$ & 10 &  & 43.1 & $0.0871\pm0.0871$ &  1 &  & 43.05 & $0.1868\pm0.1868$ & 2  \\
    43.0 & $0.1202\pm0.1202$ &  2 &  &      &                   &    &  &       &                   &    \\
    43.3 & $0.0704\pm0.0704$ &  1 &  &      &                   &    &  &       &                   &    \\
   \tableline
         &                   & 80 &  &      &                   & 35 &  &       &                   & 45 
 \enddata
\tablecomments{The uncertainties include Poisson errors, possible 
AGN contamination, and line mis-identification.
}
\end{deluxetable}

\begin{deluxetable}{lccccc}
 \tablewidth{0pt}
 \tablecaption{\label{tab:lfparam} H$\alpha$ luminosity function parameters
 for emission-line galaxies }
 \tablehead{
   \colhead{$z$ range} & \colhead{$N_{gal}$} &
   \colhead{$\phi^* (10^{-3} \mbox{Mpc}^{-3})$} &
   \colhead{$\alpha$} &
   \colhead{$\mbox{log} L^{*}$}  &
   \colhead{SFRD ($\mbox{M}_{\odot} \mbox{yr}^{-1} \mbox{Mpc}^{-3}$)}
   }
 \startdata
   \tableline
            &    & $0.788\pm0.881$ & $-1.78\pm0.46$  & $42.72\pm0.23$ & $0.088\pm0.056$ \\
   0.7--1.4 & 35 & $2.704\pm0.841$ & $-1.39$ (fixed) & $42.48\pm0.12$ & $0.094\pm0.023$ \\
            &    & $5.046\pm1.125$ & $-1.0$ (fixed)  & $42.27\pm0.09$ & $0.074\pm0.014$ \\
            &    & $0.848\pm0.289$ & $-1.8$ (fixed)  & $42.79\pm0.12$ & $0.187\pm0.033$ \\
   \tableline               
            &    & $1.745\pm1.578$ & $-1.90\pm0.39$  & $42.72\pm0.20$ & $0.265\pm0.174$ \\
   1.4--1.9 & 45 & $4.096\pm1.129$ & $-1.39$ (fixed) & $42.54\pm0.10$ & $0.164\pm0.034$ \\
            &    & $7.106\pm1.851$ & $-1.0$ (fixed)  & $42.37\pm0.09$ & $0.129\pm0.019$ \\
            &    & $1.491\pm0.495$ & $-1.8$ (fixed)  & $42.80\pm0.12$ & $0.334\pm0.063$ \\
   \tableline
            &    & $4.241\pm3.553$ & $-1.39\pm0.43$  & $42.46\pm0.19$ & $0.138\pm0.058$ \\
   0.7--1.9 & 80 & $3.303\pm0.512$ & $-1.39$ (fixed) & $42.54\pm0.06$ & $0.132\pm0.012$ \\
            &    & $5.888\pm0.514$ & $-1.0$ (fixed)  & $42.38\pm0.03$ & $0.112\pm0.009$ \\
            &    & $1.122\pm0.188$ & $-1.8$ (fixed)  & $42.82\pm0.04$ & $0.269\pm0.021$ \\
 \enddata
 \tablecomments{
 The values are calculated for $\Omega_{m}=0.3$, $\Omega_{\Lambda}=0.7$,
 and $h=71 \mbox{km} \mbox{s}^{-1} \mbox{Mpc}^{-1}$.
 The adopted form of the luminosity function is
 $\phi(L)=\phi^* (L/L^*)^{1+\alpha} \mbox{exp}(-L/L^*)$.
 The SFR density column is inferred from
 the H$\alpha$ luminosity density, using
 $SFR(H\alpha) (\mbox{M}_{\odot} \mbox{yr}^{-1}) = 7.9\times10^{-42} L(H\alpha) \mbox{erg} \mbox{s}^{-1}$
 from Kennicutt (1998).
 Here, the total luminosity density is derived using the luminosity function parameters :
 $L_{\mbox{\tiny{tot}}}=\phi^* L^* \Gamma(\alpha+2) $.
 }
\end{deluxetable}

\end{document}